\documentclass[3p, number, sort&compress, times, lefttitle]{elsarticle}

\journal{Elsevier}

\usepackage{amsmath,amsthm,amssymb,amsfonts}
\usepackage{graphicx}
\usepackage{subcaption}
\usepackage{bm}
\usepackage{siunitx}
\usepackage[dvipsnames]{xcolor}
\usepackage{bigints}
\usepackage[colorlinks=true]{hyperref}
\renewcommand{\Re}{\mathbb{R}}

\newcommand{\fref}[1]{Fig.~\ref{#1}}
\newcommand{\sref}[1]{Section~\ref{#1}}
\newcommand{\vm}[1]{\bm{#1}}
\newcommand{\vx}{\vm{x}}

\newcommand{\set}[1]{\mathbb{#1}}

\DeclareMathOperator*{\argmin}{arg\,min}

\renewcommand{\bezier}{B{\'e}zier}

\newcommand{\revone}[1]{#1}
\newcommand{\revtwo}[1]{#1}

\begin{document}

\title{Spectral extended finite element method for band structure calculations
in phononic crystals}

\author[1]{Eric B.\ Chin\corref{cor1}}
\ead{chin23@llnl.gov}

\author[2]{Amir Ashkan Mokhtari}

\author[2]{Ankit Srivastava}

\author[3]{N.\ Sukumar}

\cortext[cor1]{Corresponding author}

\address[1]{Lawrence Livermore National Laboratory, 7000 East Avenue,
Livermore, CA 94550, USA}
\address[2]{Department of Mechanical, Materials, and Aerospace Engineering
Illinois Institute of Technology, Chicago, IL 60616, 
USA}
\address[3]{Department of Civil and Environmental Engineering University of
California, Davis, CA 95616, USA}

\begin{abstract}
In this paper, we compute the band structure of one- and two-dimensional
phononic composites using the extended finite element method (X-FEM) on
structured higher-order (spectral) finite element meshes.  On using
partition-of-unity enrichment in finite element analysis, the X-FEM permits use
of structured finite element meshes that do not conform to the geometry of holes
and inclusions.  This eliminates the need for remeshing in phononic shape
optimization and topology optimization studies.  In two dimensions, we adopt
rational B{\'e}zier representation of curved (circular) geometries, and
construct suitable material enrichment functions to model two-phase composites.
A Bloch-formulation of the elastodynamic phononic eigenproblem is adopted.
Efficient computation of weak form integrals with polynomial integrands is
realized via the homogeneous numerical integration scheme---a method that uses
Euler's homogeneous function theorem and Stokes's theorem to reduce integration
to the boundary of the domain. Ghost penalty stabilization is used on finite
elements that are cut by a hole.  Band structure calculations on perforated
(circular holes, elliptical holes, and holes defined as a level set) materials
as well as on two-phase phononic crystals are presented that affirm the sound
accuracy and optimal convergence of the method on structured, higher-order
spectral finite element meshes.  Several numerical examples are presented to
demonstrate the advantages of $p$-refinement made possible by the spectral
extended finite element method.  In these examples, fourth-order spectral
extended finite elements deliver $\mathcal{O}(10^{-8})$ accuracy in frequency
calculations with more than thirty-fold fewer degrees-of-freedom when compared
to quadratic finite elements.
\end{abstract}


\maketitle

% \linenumbers

\section{Introduction}\label{sec:introduction}

The study of wave propagation through macro-scale periodic structures has
recently been a topic of significant interest.  Research on this topic has been
conducted under the umbrella of
photonics~\cite{yablonovitch1987inhibited,john1987strong},
phononics~\cite{Mokhtari2019phase,lu2009phononic,maldovan2013sound,hussein2014dynamics},
and metamaterials~\cite{liu2000locally,yu2018mechanical}, with the goal of
controlling the flow behavior of light, elastic and acoustic waves, heat, or
mechanical vibrations in suitably designed materials. Within the scope of
phononic structures, a better understanding of periodic oscillations can enable
materials capable of wave guiding and
redirection~\cite{cervera2001refractive,khelif2003trapping,aljahdali2016high,Mokhtari2020},
wave focusing and
amplification~\cite{yang2004focusing,sukhovich2008negative,zhang2009focusing,lin2009gradient,chen2014enhanced},
and wave absorption and isolation~\cite{liang2010an,mei2012dark,fleury2014sound}.
Additionally, the study of phononic structures has uncovered many interesting
phenomena and novel applications, such as ultrasonic
tunneling~\cite{yang2002ultrasound}, the inverse Doppler
effect~\cite{reed2003reversed,zhai2016inverse}, hypersonic
control~\cite{gorishnyy2005hypersonic}, light/sound
interactions~\cite{vanlaer2015interaction}, and processing of quantum
information~\cite{arrangoiz2018coupling}, to name a few.

Historically, numerical efforts in this direction have been driven towards the
calculation of the so-called band structure~\cite{kushwaha1993acoustic} of the
periodic system---a graphical representation of the frequency-wavevector pairs
that satisfy a certain kind of dispersion relationship for the system. The band
structure diagram makes characteristics of band gaps in the periodic structure
(such as direction, magnitude, phase, and whether the gap is direct or indirect)
readily apparent. Additionally, research areas such as phononic topology
optimization~\cite{sigmund2003systematic,diaz2005design,halkjaer2006maximizing,rupp2007design,bilal2011ultrawide}
and inverse problems in dynamic homogenization~\cite{nemat2011homogenization}
depend heavily on the speed, efficiency, and accuracy of the band structure
calculating algorithm.

To construct the band structure, the fundamental Bloch-periodic problem solved
in the unit phononic cell is: for a given wavevector ($k$-point) in the
irreducible Brillouin zone, find the elastodynamic eigenpairs (eigenfrequencies
and eigenmodes).  Many computational algorithms have been developed to solve
this problem.  The most common of these approaches is the plane wave expansion
(PWE)
method~\cite{Mokhtari2019prop,ho1990existence,sigalas1992elastic,Sigalas1993,kushwaha1993acoustic,kushwaha1994theory},
which expands both the material properties and the displacement field in a
global trigonometric (Fourier) basis. PWE is easy to implement, but the method
suffers from slow convergence, especially for cases with sharp material
contrasts.  Variational methods constitute another broad category of primary
phononic computational techniques.  These include the Rayleigh
quotient~\cite{bathe1976numerical} and the mixed-variational
methods~\cite{nemat1972harmonic,nemat1975harmonic,minagawa1976harmonic,srivastava2014mixed}.
Of these techniques, the mixed-variational method shows the greatest convergence
rates~\cite{babuska1978numerical} and outperforms the PWE as
well~\cite{lu2016variational}.  Another major technique is the multiple
scattering
method~\cite{kafesaki1999multiple,liu2000elastic,sigalas2005classical}, which
shows good convergence properties but is limited in its applicability to fairly
regular material inclusions in the unit cell and is cumbersome to implement. In
addition to the aforementioned methods, several secondary acceleration
techniques have been developed for phononic computations. These include
multipole-multigrid methods~\cite{chern2003large}, multi-scale finite element
methods~\cite{casadei2014multiscale}, the reduced Bloch mode expansion (RBME)
technique~\cite{hussein2009reduced}, and the Bloch Mode Synthesis
technique~\cite{krattiger2014bloch}.  Further accelerations have been proposed
through massive parallelization of the existing algorithms over distributed CPUs
and/or GPUs~\cite{srivastava2015gpu}.

In this contribution, we use the spectral extended finite element method
(X-FEM)~\cite{Legay:2005:SWA,Chin:2019:MCI} to solve the Bloch-periodic
elastodynamic eigenproblem over the unit cell.  The Galerkin finite element
approach proceeds by converting the governing boundary-value problem to an
equivalent weak form, then applying a local polynomial approximation (shape
function) to trial and test functions over a discretization (mesh) of the
Bloch-periodic domain.  The finite element method has been previously applied to
solve this
problem~\cite{veres2012complexity,hladky1991analysis,hussein2009reduced},
though low-order polynomial approximations typically used within finite elements
result in a fine discretization of the domain to accurately capture the waveform
solution of the elastodynamic problem.  This shortcoming is reduced through the
use of higher-order local polynomial approximations, which are enabled by
spectral finite elements~\cite{Karniadakis:1999:SEM}.  With spectral finite
elements, solution accuracy improves exponentially in a logarithmic plot with
polynomial order ($p$-) refinement of the approximation, as opposed to linear
improvement in a logarithmic plot with discretization ($h$-) refinement.  This
enables a more accurate solution using fewer degrees of freedom.  The advantages
of spectral finite elements in generating phononic band structures have been
recognized~\cite{Wu:2013:SVC,Shi:2016:SEM}, though the combination of spectral
finite elements and extended finite elements applied to phononic band structure
calculations is new to the best knowledge of the authors.

With a conforming mesh of the Bloch-periodic domain, the finite element method
can accurately capture solution discontinuities across material boundaries and
voids.  In comparison, approximations that do not capture discontinuities across
boundaries, such as the global Fourier basis used in PWE, suffer from reduced
rates of convergence~\cite{cao2004convergence}.  However, constructing a
conforming discretization over an arbitrary domain can be a difficult task that
requires advanced mesh generation techniques.  Furthermore, iterative design
processes, such as those used in phononic topology optimization, can require
many meshes to be generated, further increasing the burden of mesh generation.
Also, conforming meshes on curved boundaries with higher-order finite elements
require a mapping to a parent element with affine geometry.  This mapping
results in a loss of consistency, which impairs higher order polynomial
reproduction on these
elements~\cite{Fried:1973:ACC,Zienkiewicz:2000:TFE,Sevilla:2011:CHC}.  To
eliminate these issues, while maintaining accurate reproduction of domain
boundaries, we employ the X-FEM, which can be used to represent boundaries
implicitly, through use of enrichment functions~\cite{Moes:1999:AFE}.  Zhao et
al.~\cite{Zhao:2015:PBS} have demonstrated the promise of the X-FEM using linear
elements to construct the band structure diagram for metamaterials. Herein, we
extend this idea to incorporate higher-order, spectral shape functions with the
X-FEM.  Use of the X-FEM with spectral basis functions introduces challenges
with respect to boundary representation, interface enrichment, and numerical
integration. Chin and Sukumar~\cite{Chin:2019:MCI} discuss methods of resolving
these challenges, and we apply these improvements to the procedures used in this
paper.

The structure of the remainder of this paper follows. In
\sref{sec:phononics-problem}, we introduce the governing elastodynamic boundary
value problem and use Bloch-Floquet theorem to transform the problem to the
phononic unit cell, where the strong form and the weak form of the problem are
derived.  In \sref{sec:spectral-fe}, higher order polynomial approximations are
introduced as the spectral finite element method.  In \sref{sec:xfem}, enriched
finite element spaces are introduced to capture hole and interface geometry in
the unit cell.  Also, the finite element equations to solve the phononic unit
cell problem are introduced.  \sref{sec:curved-geometries} describes how
rational \bezier{} curves and level set functions can be used with the extended
finite element method to represent curved geometry.  Implementation details,
including numerical integration and interface stabilization are discussed in
\sref{sec:xfem-implementation}.  In \sref{sec:results}, several numerical
examples in one dimension and two dimensions are discussed which demonstrate the
capabilities of the methodology.  Finally, in \sref{sec:conclusion}, we
summarize the developments in this paper.

\section{Phononic unit cell problem}\label{sec:phononics-problem}

\begin{figure}
  \centering
  \includegraphics[scale=0.969]{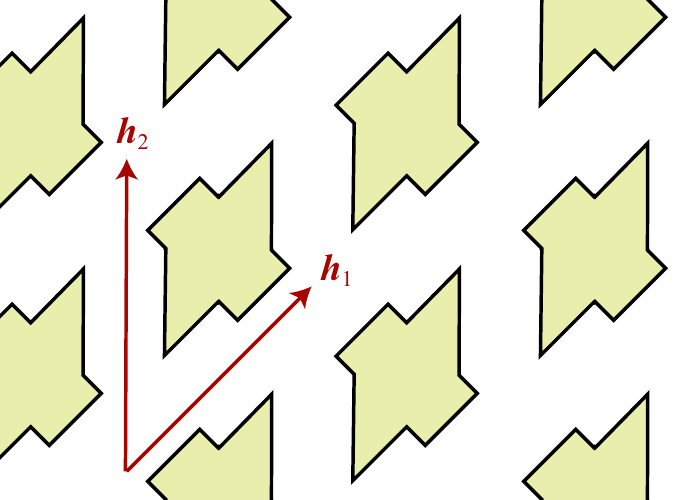}
  \caption{A two-dimensional periodic composite with basis vectors $\vm{h}_1$ 
    and $\vm{h}_2$.}\label{fig:unit-cell-basis}
\end{figure}

Consider a two-dimensional periodic composite structure whose unit cell is a
parallelogram given by $P \subset \Re^2$.  We define two basis vectors,
$\vm{h}_1$ and $\vm{h}_2$, such that $\vx = \sum_{i=1}^2 H_i \vm{h}_i$ for $H_i
\in [0, 1]$ uniquely maps to a point in $P$.  An example is illustrated in
\fref{fig:unit-cell-basis}.  The basis vectors can be used to define spatial
variation in material properties $\vm{C} (\vx)$ (the rank-four modulus tensor)
and $\rho (\vx)$ (density), since
\begin{equation}
  \vm{C} (\vx + n_i \vm{h}_i) = \vm{C} (\vx) , \qquad
  \rho (\vx + n_i \vm{h}_i) = \rho (\vx) ,
\end{equation}
for $n_i \in \mathbb{Z}$.  We define reciprocal basis vectors
\begin{equation}
  \vm{q}_1 = 2 \pi \frac{\vm{h}_2 \times \vm{e}_3}
    {\vm{h}_1 \cdot \left( \vm{h}_2 \times \vm{e}_3 \right)} , \qquad
  \vm{q}_2 = 2 \pi \frac{\vm{e}_3 \times \vm{h}_1}
    {\vm{h}_2 \cdot \left( \vm{e}_3 \times \vm{h}_1 \right)} ,
\end{equation}
where $\vm{e}_3$ is a unit vector in the (Cartesian) $z$-direction, such that
the property $\vm{h}_i \cdot \vm{q}_j = 2 \pi \delta_{ij}$ holds.  These
reciprocal basis vectors are used to define a wave vector $\vm{k}$ in a periodic
composite via $\vm{k} = \sum_{i=1}^2 Q_i \vm{q}_i$ where $Q_i \in [0,1]$.

\subsection{Strong formulation}\label{ssec:strong-form}
\begin{figure}
  \centering
  \includegraphics[scale=0.969]{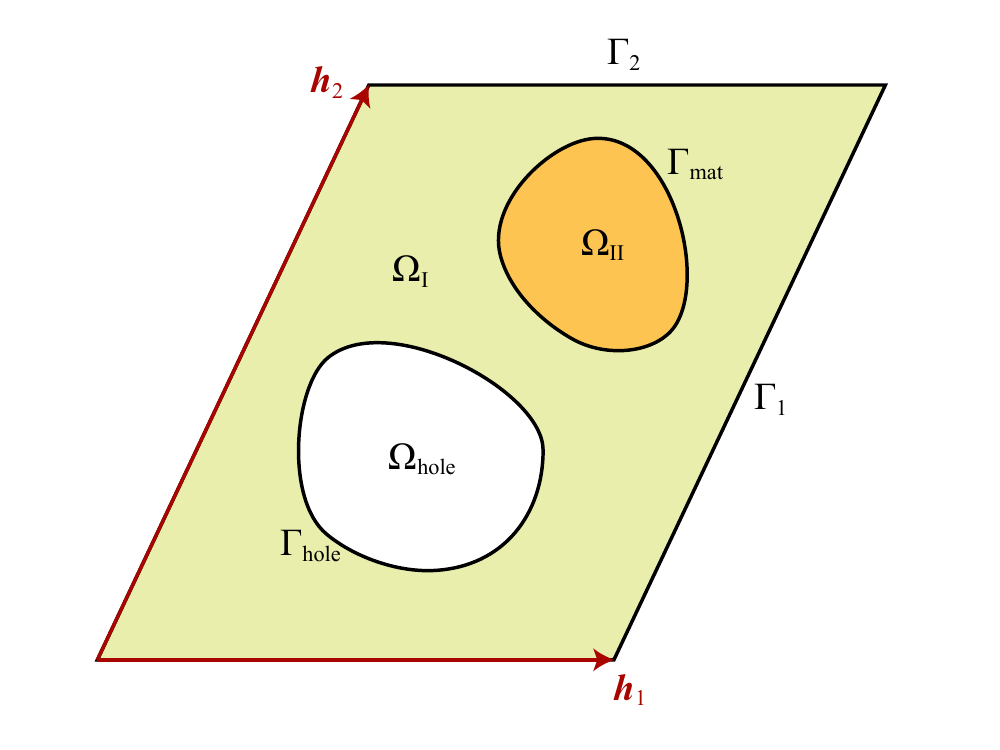}
  \caption{Bloch-periodic boundary-value problem with two distinct linear
  elastic materials and a hole.}\label{fig:unit-cell-bvp}
\end{figure}

The behavior of waves in solids is governed by the elastodynamic boundary-value
problem.  Over $\Omega \subseteq P$, the domain of the parallelogram $P$ not
including voids, the governing field equations are:
\begin{subequations}\label{eq:2d-governing}
\begin{align}
  \nabla \cdot \vm{\sigma} & = \rho \ddot{\vm{u}} 
    & \text{(balance of linear momentum)} , \\
  \vm{\sigma} & = \vm{\sigma}^T
    & \text{(balance of angular momentum)} , \\
  \vm{\sigma} & = \vm{C} : \vm{\varepsilon} 
    & \text{(constitutive law)} , \\
  \vm{\varepsilon} & = \frac{1}{2} \left( \nabla \vm{u} 
      + ( \nabla \vm{u} )^T \right)
    & \text{(small-strain kinematics)} ,
\end{align}
\end{subequations}
where $\vm{\sigma} := \vm{\sigma} (\vx, t)$ is the Cauchy stress tensor, $\rho :=
\rho (\vx)$ is the scalar density field, $\vm{u} := \vm{u} (\vx, t)$ is the
displacement field, and $\vm{\varepsilon} := \vm{\varepsilon} (\vx, t)$ is the
strain tensor.  We assume plane strain conditions ($\varepsilon_{13} =
\varepsilon_{23} = \varepsilon_{33} = 0$).  Since the problem is time-harmonic
with angular frequency $\omega$, the field quantities can be decomposed as
follows:
\begin{equation}\label{eq:time-harmonic-fields}
  \vm{u} (\vx, t) = \vm{u} (\vx) \exp (-i \omega t) , \qquad 
  \vm{\sigma} (\vx, t) = \vm{\sigma} (\vx) \exp (-i \omega t) , \qquad 
  \vm{\varepsilon} (\vx, t) = \vm{\varepsilon} (\vx) \exp (-i \omega t) .
\end{equation}
Combining \eqref{eq:2d-governing} and \eqref{eq:time-harmonic-fields}, we
recover the eigenproblem
\begin{equation}
  \nabla \cdot (\vm{C} (\vx) : \nabla_s \vm{u} (\vx)) 
  = -\lambda \rho (\vx) \vm{u} (\vx) ,
\end{equation}
where $\lambda = \omega^2$ is the eigenvalue and $\nabla_s (\cdot)$ denotes the
symmetric gradient.  According to the Bloch-Floquet theorem, the displacement
and stress fields across unit cells are related by
\begin{equation}
  \vm{u} (\vx + \vm{h}_i) 
    = \vm{u} (\vx) \exp \bigl( i \vm{k} \cdot \vm{h}_i \bigr) , \qquad
  \vm{\sigma} (\vx + \vm{h}_i) 
    = \vm{\sigma} (\vx) \exp \bigl( i \vm{k} \cdot \vm{h}_i \bigr) .
\end{equation}
These relationships can be used to develop Bloch-periodic boundary conditions on
the unit cell, which complete the strong form of the unit cell problem.  The
problem can be stated as follows:
\begin{subequations}\label{eq:2d-strong-form}
\begin{alignat}{2}
  \nabla \cdot (\vm{C} (\vx) : \nabla_s \vm{u} (\vx))
    + \lambda \rho (\vx) \vm{u} (\vx) & = 0 \quad
    && \text{in } \Omega , \label{eq:2d-pde}\\
  \vm{u} (\vx + \vm{h}_i)
    & = \vm{u} (\vx) \exp (i \vm{k} \cdot \vm{h}_i) \quad
    && \text{on } \Gamma_i \text{ for } i = 1, 2 , \label{eq:periodic-dirichlet}\\
  \vm{\sigma} (\vx + \vm{h}_i) \cdot \vm{n} (\vx + \vm{h}_i)
    & = \vm{\sigma} (\vx) \cdot \vm{n} (\vx) \exp (i \vm{k} \cdot \vm{h}_i) \quad
    && \text{on } \Gamma_i \text{ for } i = 1, 2 ,
\end{alignat}
\end{subequations}
where $\vm{n}$ is the unit normal vector pointing outward from the domain, and
$\Gamma_1$ and $\Gamma_2$ are boundary regions that align with basis vectors
$\vm{h}_2$ and $\vm{h}_1$, respectively. These quantities are illustrated in
\fref{fig:unit-cell-bvp}. Additionally, within the unit cell, we assume voids
and/or two distinct, linear elastic materials are present.  Holes are located in
the domain $\Omega_\text{hole}$, with boundary $\Gamma_\text{hole}$, and the
domain of material $m$ is $\Omega_m$ for $m = I, II$.  The boundary between the
two materials is $\Gamma_\text{mat}$. With these additional assumptions, we can
explicitly state the boundary and interface conditions that must be satisfied
within the unit cell:
\begin{subequations}
\begin{alignat}{2}
  \vm{\sigma} (\vx) \cdot \vm{n} (\vx)
    & = \vm{0} \quad
    && \text{on } \Gamma_\text{hole} , \label{eq:hole-dirichlet}\\
  [[ \vm{u} (\vx) ]]
    & = \vm{0} \quad
    && \text{on } \Gamma_\text{mat} , \label{eq:interface-dirichlet}\\
  [[ \vm{\sigma} (\vx) \cdot \vm{n} (\vx) ]]
    & = \vm{0} \quad
    && \text{on } \Gamma_\text{mat} , \label{eq:interface-traction}
\end{alignat}
\end{subequations}
where $[[\cdot]]$ is the jump operator that represents the jump in its argument
across the interface.  These conditions along with \eqref{eq:2d-strong-form}
constitute the strong form of the Bloch-periodic boundary-value problem.

In the developments that follow, we will also consider one-dimensional periodic
domains $\Omega = (0, a)$ composed of two, distinct linear elastic materials
with properties $(E_1, \rho_1)$ and $(E_2, \rho_2)$ and an interface at $x = h$.
For this problem, the strong form reduces to
\begin{equation}
\begin{gathered}
  ( E(x) u^\prime (x)) ^\prime + \lambda \rho (x) u (x) 
    = 0 \qquad \textrm{ in } \Omega = (0,a) , \\
  u(a) = \exp (i ka) u(0),  \qquad
  E_2u^\prime(a) = \exp (i ka) E_1 u^\prime(0) , \\
  E_1 u^\prime(h^-) = E_2 u^\prime(h^+) , \qquad
  u (h^-) = u (h^+) ,
\end{gathered}
\end{equation}
where $x = h^-$ and $x = h^+$ are the locations just to the left and right of
the interface, respectively.
\subsection{Bloch-periodic weak formulation}\label{ssec:bloch-weak-form}
The strong form of the Bloch-periodic boundary-value problem can be stated
equivalently in a weak form which is amenable to computations using the finite
element method~\cite{Sukumar:2009:CEF}.  Let the trial displacement function
$\vm{u} (\vx) : \Omega \rightarrow \mathbb{C}$ and test function $\vm{v} (\vx) :
\Omega \rightarrow \mathbb{C}$ be in $H^1 (\Omega)$. Following the standard
Galerkin procedure, we multiply \eqref{eq:2d-pde} by $\vm{v}^* (\vx)$, the
complex conjugate of $\vm{v} (\vx)$, and integrate over $\Omega$ to obtain
\begin{equation}
  \int_\Omega \vm{v}^* (\vx) \cdot \left[ 
    \nabla \cdot (\vm{C} (\vx) : \nabla_s \vm{u} (\vx))
    + \lambda \rho (\vx) \vm{u} (\vx) \right] d\vx = 0 .
\end{equation}
Owing to the discontinuous material density and modulus tensor, the integral can
be decomposed to
\begin{equation}
  \int_{\Omega_I} \vm{v}^* (\vx) \cdot \left[ 
    \nabla \cdot (\vm{C}_I : \nabla_s \vm{u} (\vx))
    + \lambda \rho_I \vm{u} (\vx) \right] d\vx
  + \int_{\Omega_{II}} \vm{v}^* (\vx) \cdot \left[ 
    \nabla \cdot (\vm{C}_{II} : \nabla_s \vm{u} (\vx))
    + \lambda \rho_{II} \vm{u} (\vx) \right] d\vx = 0 ,
\end{equation}
where $(\vm{C}_I, \rho_I)$ and $(\vm{C}_{II}, \rho_{II})$ represent material
properties in $\Omega_I$ and $\Omega_{II}$, respectively.  Integration by parts,
using the divergence theorem on the first term of each integral, the symmetry of
$\vm{\sigma} (\vx)$, and the boundary condition \eqref{eq:hole-dirichlet} yields
\begin{multline}\label{eq:weak-form-3}
  - \int_{\Omega} \nabla_s \vm{v}^* (\vx) : \vm{C} (\vx)
    : \nabla_s \vm{u} (\vx) \, d\vx 
  + \sum_{i=1}^2 \int_{\Gamma_i} 
    \left[ \vm{v}^* (\vx) \cdot \vm{\sigma} (\vx) \cdot \vm{n} (\vx)
    + \vm{v}^* (\vx + \vm{h}_i) \cdot \vm{\sigma} (\vx + \vm{h}_i) 
      \cdot \vm{n} (\vx + \vm{h}_i) \right] dS \\
  + \int_{\Gamma_\text{mat}} 
    [[ \vm{v}^* (\vx) \cdot \vm{\sigma} (\vx) \cdot \vm{n} (\vx) ]] \, dS
  + \lambda \int_{\Omega} 
    \rho (\vx) \vm{v}^* (\vx) \cdot \vm{u} (\vx) \, d\vm{x} = 0 .
\end{multline}
To simplify \eqref{eq:weak-form-3} further, we require $\vm{v} (\vx)$ satisfy
conditions \eqref{eq:interface-dirichlet} and \eqref{eq:periodic-dirichlet}.
These conditions give
\begin{multline}\label{eq:weak-form-4}
  - \int_{\Omega} \nabla_s \vm{v}^* (\vx) : \vm{C} (\vx)
    : \nabla_s \vm{u} (\vx) \, d\vx 
  + \sum_{i=1}^2 \int_{\Gamma_i} 
    \vm{v}^* (\vx) \cdot \left[ \vm{\sigma} (\vx) \cdot \vm{n} (\vx)
    + \exp(- i \vm{k} \cdot \vm{h}_i) \vm{\sigma} (\vx + \vm{h}_i) 
      \cdot \vm{n} (\vx + \vm{h}_i) \right] dS \\
  + \lambda \int_{\Omega} 
    \rho (\vx) \vm{v}^* (\vx) \cdot \vm{u} (\vx) \, d\vm{x} = 0 .
\end{multline}
Now we apply boundary condition \eqref{eq:interface-traction} to arrive at
\begin{equation}
  - \int_{\Omega} \nabla_s \vm{v}^* (\vx) : \vm{C} (\vx)
    : \nabla_s \vm{u} (\vx) \, d\vx
  + \lambda \int_{\Omega} 
    \rho (\vx) \vm{v}^* (\vx) \cdot \vm{u} (\vx) \, d\vm{x} = 0 .
\end{equation}

Stated formally, the weak form is: given $\vm{C} (\vx)$, $\rho (\vx)$, and
$\vm{k} \in \Re^2$, find $\vm{u} \in \mathcal{S}$ and $\lambda := \omega^2 \in
\Re_+$, such that
\begin{subequations}
\begin{align}\label{eq:weak-form-final}
  a(\vm{u}, \vm{v}) + \frac{\gamma_K}{h^2} j(\vm{u}, \vm{v}) = \lambda 
    [ & b(\vm{u}, \vm{v}) + \gamma_M j(\vm{u}, \vm{v}) ] \quad
    \forall \vm{v} \in \mathcal{S}
  \intertext{where}
  a ( \vm{u}, \vm{v} ) := \int_\Omega \nabla_s \vm{v}^* (\vx) : \vm{C} (\vx)
    : \nabla_s \vm{u} (\vx) \, d \vx , & \quad
  b ( \vm{u}, \vm{v} ) := \int_\Omega \rho (\vx) \vm{v}^* (\vx) 
    \cdot \vm{u} (\vx) \, d \vx , \\ \label{eq:fe-space}
  \mathcal{S} := \Bigl\{ \;
    \vm{u} \in \left[ H^1 (\Omega_I \cup \Omega_{II}) \right]^2 \; : \;
    [[ \vm{u} ]] = \vm{0} \text{ on } \Gamma_\text{mat},
    & \,\,\,
    \vm{u} (\vx + \vm{h}_i) = \vm{u} (\vx) \exp(i \vm{k} \cdot \vm{h}_i)
    \text{ on } \Gamma_i \,\, (i = 1, 2) \; \Bigr\} .
\end{align}
\end{subequations}
In~\eqref{eq:weak-form-final}, the term $j ( \vm{u}, \vm{v} )$ is a
\textit{ghost penalty}~\cite{Burman:2010:GP} stabilization term to improve
matrix-conditioning, $\gamma_K$ and $\gamma_M$ are constants that control the
amount of stabilization energy, and $h$ is a characteristic length defined at
the element level (see \sref{ssec:interface-stabilization}).  The ghost penalty
term is defined in \sref{ssec:interface-stabilization}.

Reduced to one dimension, the weak form is: given $E (\vx)$, $\rho (\vx)$, and
$k \in \Re$, find $u \in {\cal S}$ and $\lambda := \omega^2 \in \Re_+$ such
that
\begin{subequations}
\begin{align}
  a(v,u) =  \lambda \, b(v,u) \quad \forall v \in {\cal S}, \quad & 
  \intertext{where}
  a(v,u) = \int_0^a v^{*\prime} (x) E (x) u^\prime (x) \, dx , \quad 
  b(v, u) = & \int_0^a \rho(x) v^* (x) u (x) \, dx , \\
  {\cal S} = \Bigl\{ \;
    u \in H^1 (0,a) \, : \,
    u (h^-) = u (h^+), \ \  u (a) &\ = u (0) \exp(i k a) \; \Bigr\} .
\end{align}
\end{subequations}

\section{Spectral finite element method}\label{sec:spectral-fe}

With X-FEM, geometric features such as holes and material interfaces are not
represented by the mesh, so a simple finite element mesh suffices. For
applications in solid continua, quadrilateral finite elements are preferred to
triangular finite elements, and discretizations aligned with the unit cell basis
vectors $\vm{h}_1$ and $\vm{h}_2$ provide numerous benefits and simplifications
as well. From the perspective of the X-FEM, additional advantages include easier
implementation for higher-order elements, compatibility with the homogeneous
numerical integration (HNI) scheme (requires basis functions that are a linear
combination of homogeneous polynomials; see~\sref{ssec:numerical-integration}
for details), and simplified computation of the ghost penalty stiffness term
introduced in~\sref{ssec:interface-stabilization} (over rectangular domains).
Accordingly, we restrict ourselves to finite element meshes aligned with the
unit cell basis in the subsequent developments.

\begin{figure}
  \centering
  \begin{subfigure}{1.575in}
    \centering
    \includegraphics[scale=0.969]{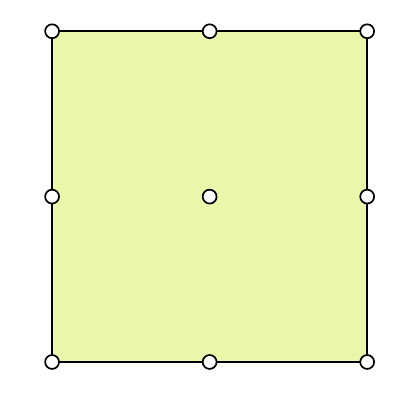}
    \caption{}\label{fig:spectral-nodes-p2}
  \end{subfigure}
  \begin{subfigure}{1.575in}
    \centering
    \includegraphics[scale=0.969]{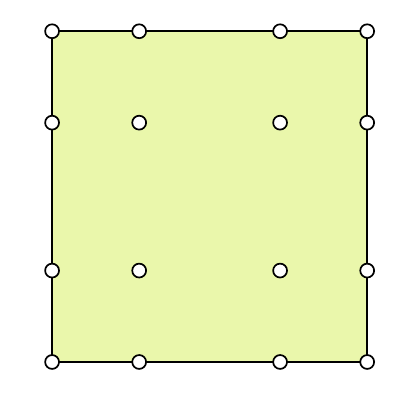}
    \caption{}\label{fig:spectral-nodes-p3}
  \end{subfigure}
  \begin{subfigure}{1.575in}
    \centering
    \includegraphics[scale=0.969]{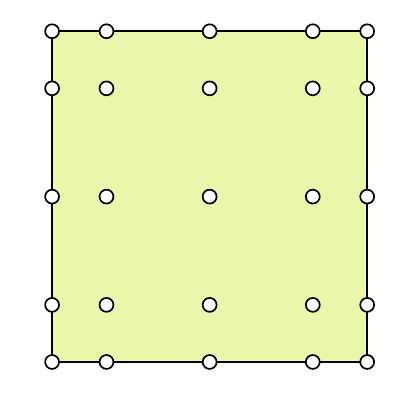}
    \caption{}\label{fig:spectral-nodes-p4}
  \end{subfigure}
  \begin{subfigure}{1.575in}
    \centering
    \includegraphics[scale=0.969]{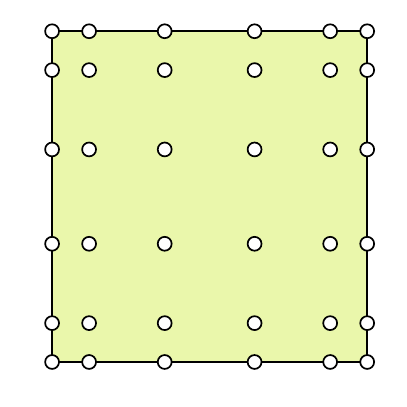}
    \caption{}\label{fig:spectral-nodes-p5}
  \end{subfigure}
  \caption{Gauss-Lobatto nodes for $p$-th order spectral elements.
           (a) $p = 2$, (b) $p = 3$, (c) $p = 4$, and (d) $p = 5$.}
  \label{fig:spectral-nodes}
\end{figure}

Consider a tessellation $\mathcal{T}$ of the parallelogram domain $\overline{P}
= \overline{\Omega \cup \Omega_\text{hole}}$ into a mesh of $M$ elements.  The
domain of the $e$-th element of the mesh is $\Omega_e \subset \Re^2$.  On
$\Omega_e$, we desire reproduction of $p$-th order polynomials, which is
accomplished through a Lagrange interpolation of $(p+1)^2$ nodal values.  We
first develop this interpolant over a parent element, $\Xi = [-1,1]^2$, then map
it to the domain $\Omega_e$.  Node locations $\vm{\xi}_I \in [-1, 1]^2$
$\bigl($for $I = 1, \dotsc , (p+1)^2 \bigr)$ are chosen as the components of the
tensor product of the Gauss-Lobatto points, $\{ \xi_1, \dotsc, \xi_{p+1} \}$ and
$\{ \eta_1, \dotsc, \eta_{p+1} \}$.  This choice of nodal locations limits
oscillation in the interpolation of smooth data (Runge phenomenon). The
placement of nodes at the Gauss-Lobatto locations for spectral elements of order
$p = 2$ to $p = 5$ is illustrated in~\fref{fig:spectral-nodes}.  At each node we
define a Lagrange shape function,
\begin{equation}\label{eq:shape-function}
  N_{I(i,j)} (\vm{\xi}) = \left(
      \prod_{\substack{k=1 \\ k \neq i}}^{p+1} 
        \frac{\xi - \xi_k}{\xi_i - \xi_k}
    \right) \left(
      \prod_{\substack{\ell=1 \\ \ell \neq j}}^{p+1}
        \frac{\eta - \eta_\ell}{\eta_j - \eta_\ell}
    \right) \quad \forall \vm{\xi} \in [-1, 1]^2,
\end{equation}
where $I(i,j) : \left\{ 1, 2, \dotsc, p+1 \right\}^2 \rightarrow \left\{1, 2,
\dotsc, (p+1)^2 \right\}$ uniquely maps components $i$ and $j$ to vector $I$.
The shape functions form a basis that uniquely interpolates polynomials up to
degree $p$ over $[-1,1]^2$.  The shape functions are interpolating since the
Kronecker delta property, $N_i (\vm{\xi}_j) = \delta_{ij}$, holds.

\begin{figure}[t]
  \centering
  \includegraphics[scale=0.969]{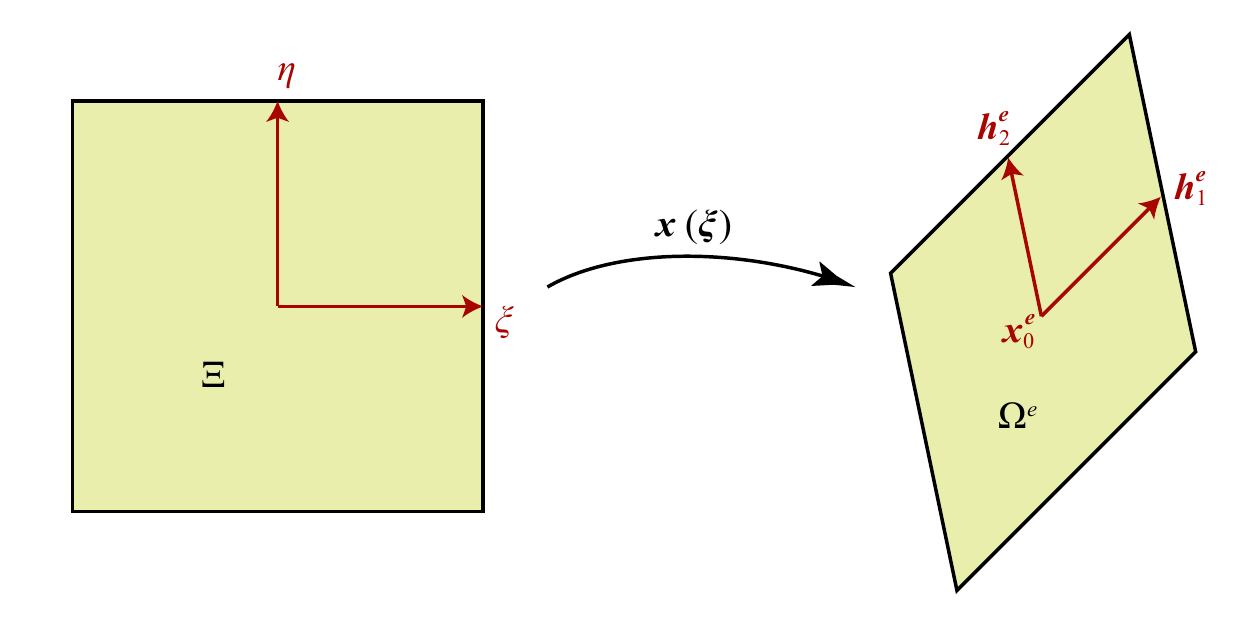}
  \caption{The isoparametric mapping between the parent domain, $\Xi$, and the
  element domain, $\Omega_e$.}\label{fig:isoparametric-map}
\end{figure}

Nodes are located in $\Omega_e$ through an invertible, affine isoparametric
mapping $\vx (\vm{\xi}) : \Xi \rightarrow \Omega_e$, which is
\begin{equation}
  \vx (\vm{\xi}) = \vm{H}^e \cdot \vm{\xi} + \vx^e_0 ,
\end{equation}
where $\vx^e_0$ denotes the centroid of the element and the columns of
$\vm{H}^e$ are the basis vectors of the element domain $\Omega_e$ ($\vm{h}^e_1$
and $\vm{h}^e_2$).  An example isoparametric mapping is shown in
\fref{fig:isoparametric-map}.  Using the shape functions, we can compute a
degree $p$ polynomial approximation of the displacement field over the element:
\begin{equation}\label{eq:uh}
  \vm{u}_e^h (\vx) = \sum_{i \in \set{I}_e} N_i (\vx) \vm{u}_i 
  \quad \forall \vx \in \Omega_e ,
\end{equation}
where $\set{I}_e = \left\{ 1, 2, \dotsc, (p+1)^2 \right\}$ is the index set of
nodes in $\Omega_e$ and $\vm{u}_i := \vm{u} (\vx_i)$ are nodal values of
displacement. This element-level approximation is utilized in the development of
the finite element equations in \sref{sec:xfem-implementation}.

\section{Extended finite element method}\label{sec:xfem}

The extended finite element method is an instance of the partition-of-unity
finite element method~\cite{Melenk:1996:PUF}, which provides a means to include
known solution characteristics in the approximation space.  This is accomplished
by augmenting the standard finite element space with the product of
partition-of-unity functions and enrichment functions.  \revtwo{For proofs of
convergence with the partition-of-unity finite element method (and, by
extension, the X-FEM), we direct the reader to Babu\v{s}ka and
Melenk~\cite{Babuska:1997:PUM} and Babu\v{s}ka et al.~\cite{Babuska:2017:SSG}.}
Even though the X-FEM can be used to model a domain that contains both holes and
material interfaces, to simplify the exposition, we treat each problem
separately.  In the sections that follow, we detail the specific enrichment
functions used to model holes and material interfaces, respectively.
\subsection{Modeling of holes}\label{ssec:modeling-holes}
\begin{figure}
  \centering
  \begin{subfigure}{3.15in}
    \centering
    \includegraphics[scale=0.969]{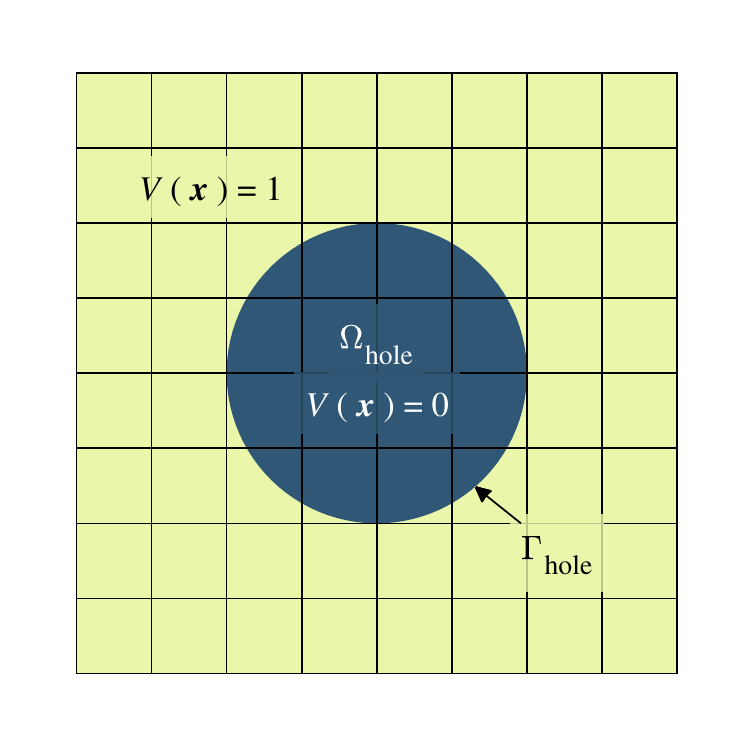}
    \caption{}\label{fig:hole-enrichment}
  \end{subfigure}
  \begin{subfigure}{3.15in}
    \centering
    \includegraphics[scale=0.969]{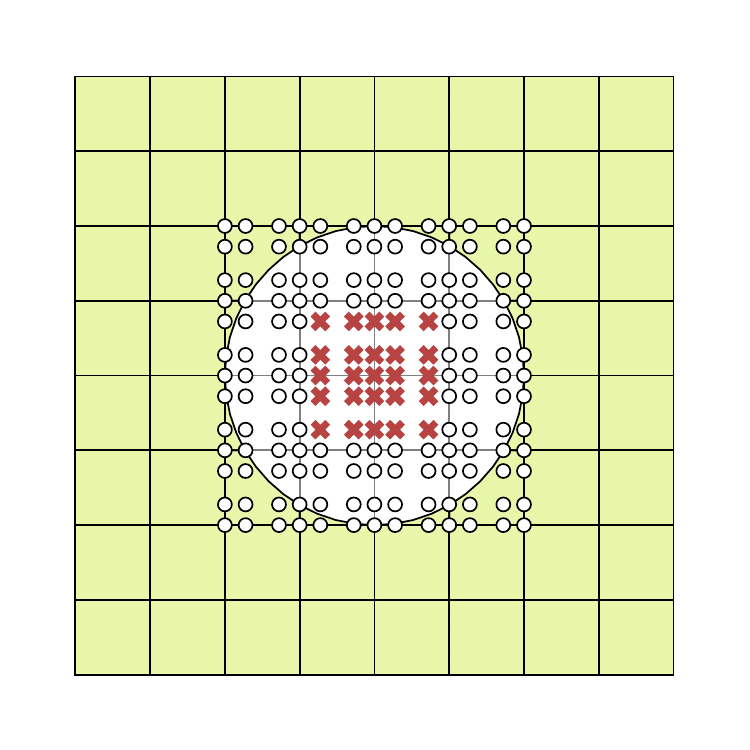}
    \caption{}\label{fig:hole-enrichment-dofs}
  \end{subfigure}
  \caption{Modeling a void with the X-FEM. (a) Enrichment function for a
           circular void, and (b) enriched DOFs (circles) and removed DOFs
           (crosses) in the vicinity of a void for bilinear finite elements.}
  \label{fig:hole-enrichment-and-dofs}
\end{figure}

To capture the effect of holes on the finite element approximation, we construct
an enrichment function based on the void geometry as
follows~\cite{Daux:2000:ABI,Sukumar:2001:MHI}:
\begin{equation}\label{eq:V}
  V (\vx) = \begin{cases}
    1 & \textrm{if } \vx \notin \Omega_\text{hole} \\
    0 & \textrm{if } \vx \in \Omega_\text{hole}
  \end{cases} \quad \forall \vx \in \Omega.
\end{equation}
An example of an enrichment function for a circular hole is illustrated in
\fref{fig:hole-enrichment}.  For an element in which $\Omega_e \cap
\Omega_\text{hole} \neq \emptyset$, the extended finite element approximation of
the displacement field $\vm{u} (\vx) : \overline{\Omega} \rightarrow \Re^2$
restricted to $\overline{\Omega_e} \backslash \Omega_\text{hole}$
is~\cite{Daux:2000:ABI}:
\begin{equation}\label{eq:uh-hole}
  \vm{u}_e^h(\vx) = \sum_{i \in \set{I}_e} N_i(\vx) V(\vx) \vm{u}_i ,
\end{equation}
where $\vm{u}_i$ are the nodal displacement degrees of freedom \revtwo{and
$N_i(\vx)$ are the spectral finite element basis functions introduced
in~\eqref{eq:shape-function}}.  Based on the location of $\Omega_\text{hole}$
with respect to $\Omega_e$, two cases are possible:
\begin{enumerate}
  \item $\Omega_e \cap \Gamma_\text{hole} = \emptyset$: the entirety of
    $\Omega_e$ is located in $\Omega_\text{hole}$.
  \item $\Omega_e \cap \Gamma_\text{hole} \neq \emptyset$: a portion of
    $\Omega_e$ is located in $\Omega_\text{hole}$.
\end{enumerate}
For case 1, the entire element domain is not in the domain of $\vm{u} (\vx)$.
For case 2, only a portion of the element domain is in the domain of $\vm{u}
(\vx)$, and the nodal degrees of freedom in this element should only affect the
portion of the element not in $\Omega_\text{hole}$.  This is accomplished
through the enrichment function $V(\vx)$, which introduces a strong
discontinuity at $\Gamma_\text{hole}$.  Note that if case 1 occurs, there may
exist nodes that have no finite element basis function support.  An illustration
of this scenario is shown in~\fref{fig:hole-enrichment-dofs}.  The degrees of
freedom (DOFs) at these nodes (the ones that contain crosses
in~\fref{fig:hole-enrichment-dofs}) are removed during the solution procedure.
This effectively constrains the displacement of these nodes to zero.  This
ensures that the stiffness matrix has full rank since there is no strain energy
associated with the deformation of these nodes.

\subsection{Modeling material interfaces}\label{ssec:modeling-interfaces}
\begin{figure}
  \centering
  \includegraphics[scale=0.969]{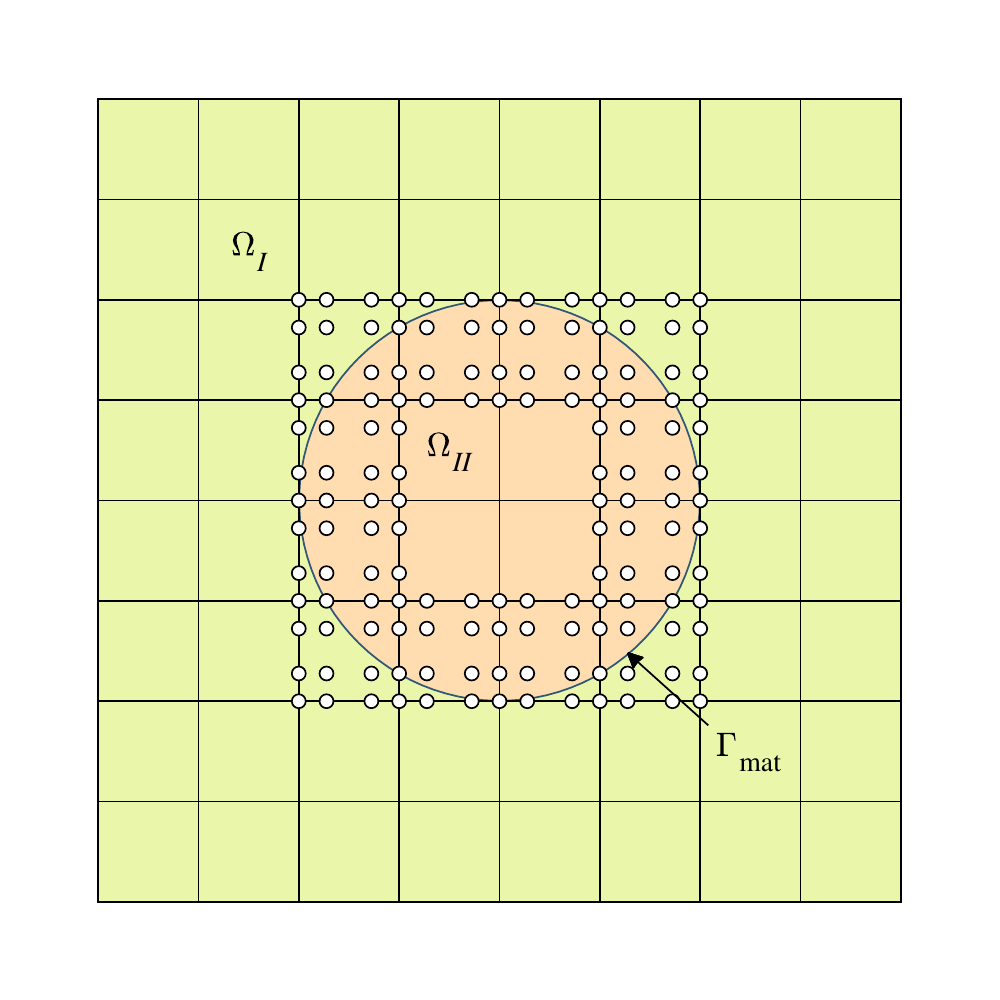}
  \caption{Two-phase composite material.  Enriched nodes are marked with 
           an open circle.}
  \label{fig:xfem-composites}
\end{figure}

In an element that intersects the material interface, i.e., $\Omega_e \cap
\Gamma_\text{mat} \neq \emptyset$, the extended finite element approximation
takes the form~\cite{Sukumar:2001:MHI}:
\begin{equation}\label{eq:uh-bimat}
  \vm{u}_e^h(\vx) = \underset{\textrm{standard FE}} {
    \underbrace{ \sum_{i \in \set{I}_e} N_i(\vx)\vm{u}_i }} +
    \underset{\textrm{enriched contribution}} {
    \underbrace{ \sum_{j \in \set{J}_e \subseteq \set{I}_e} N_j(\vx) \psi (\vx) 
    \vm{a}_{j} }} ,
\end{equation}
where $\psi (\vx)$ is the interface enrichment function, $\set{I}_e$ is the
index set of nodes in $\Omega_e$, and $\set{J}_e$ is the index set of nodes in
$\Omega_e$ whose basis function support intersects $\Gamma_\text{mat}$.  A
schematic of a two-phase composite (circular interface) that is modeled using a
quadratic finite element mesh is depicted in~\fref{fig:xfem-composites}.  In
\fref{fig:xfem-composites}, all the enriched nodes are shown as open circles.

\begin{figure}
  \centering
  \includegraphics[scale=0.969]{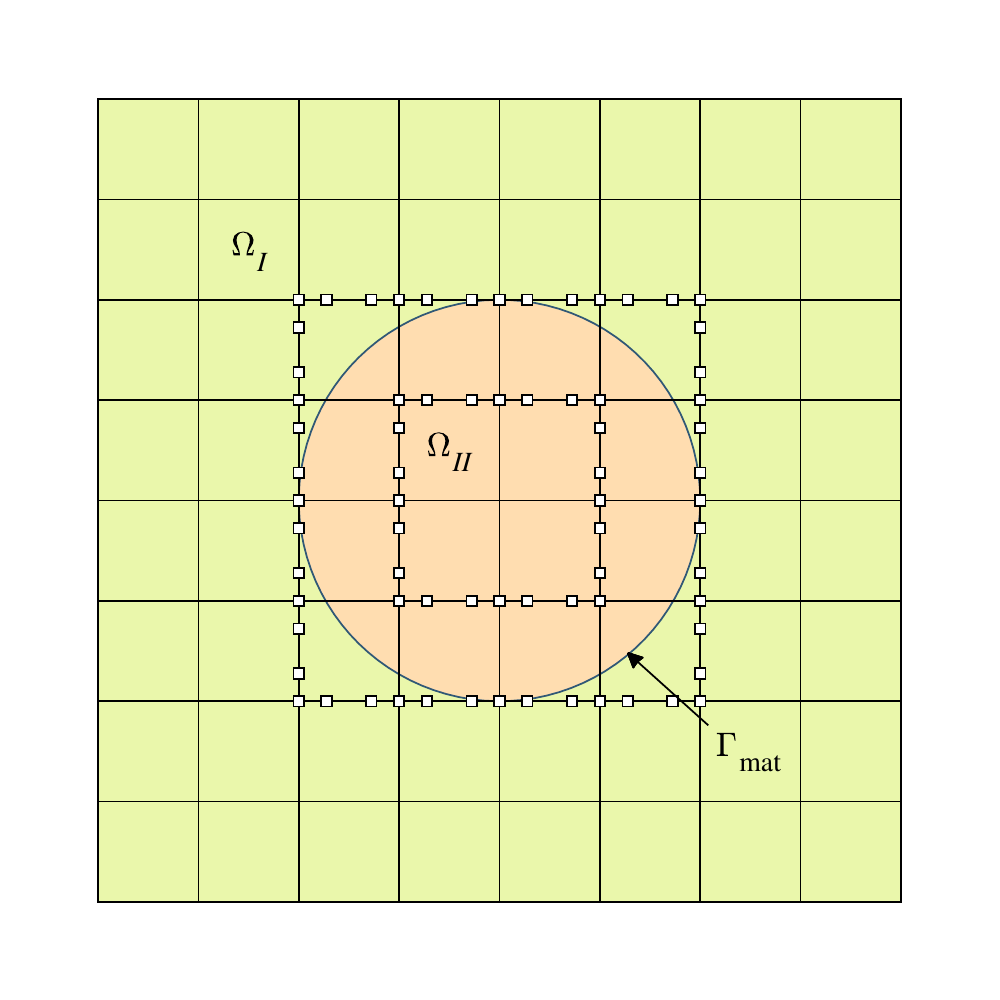}
  \caption{Nodes in $\set{I}_\text{zero}$ (marked by open squares)
           on a $8 \times 8$ cubic finite element mesh.}
  \label{fig:set-izero}
\end{figure}

\begin{figure}[t]
  \centering
  \includegraphics[scale=0.969]{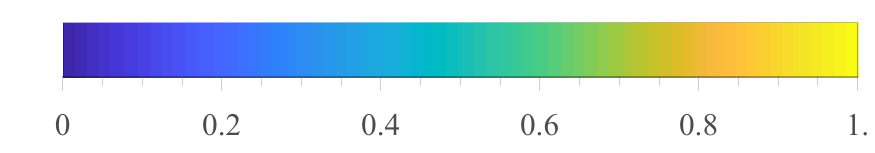} \\
  \begin{subfigure}{2.1in}
    \centering
    \includegraphics[width=\textwidth]{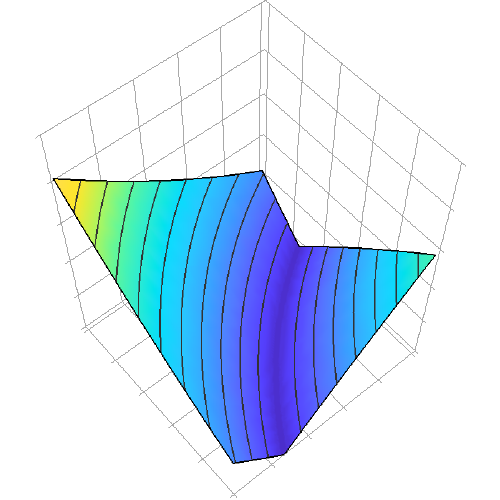}
    \includegraphics[width=\textwidth]{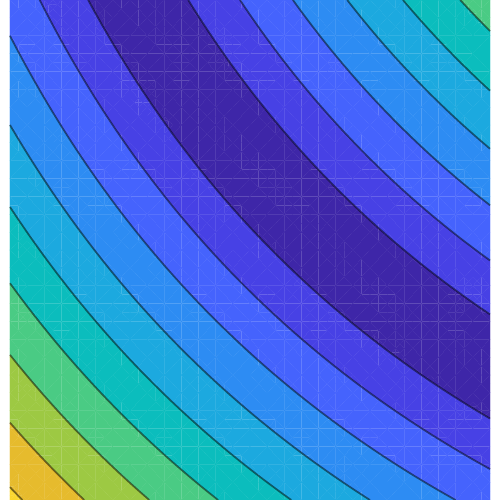}
    \caption{$| \varphi^h (\vx) |$}
    \label{fig:bimat-phih-abs}
  \end{subfigure}
  \begin{subfigure}{2.1in}
    \centering
    \includegraphics[width=\textwidth]{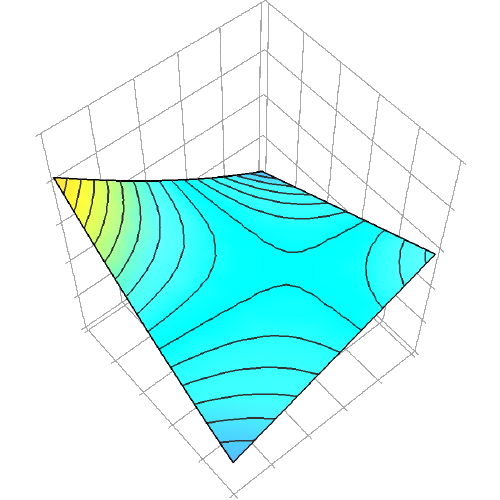}
    \includegraphics[width=\textwidth]{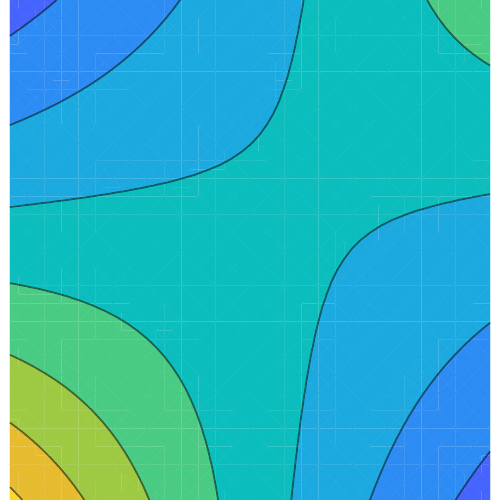}
    \caption{$g (\vx)$}
    \label{fig:bimat-g}
  \end{subfigure}
  \begin{subfigure}{2.1in}
    \centering
    \includegraphics[width=\textwidth]{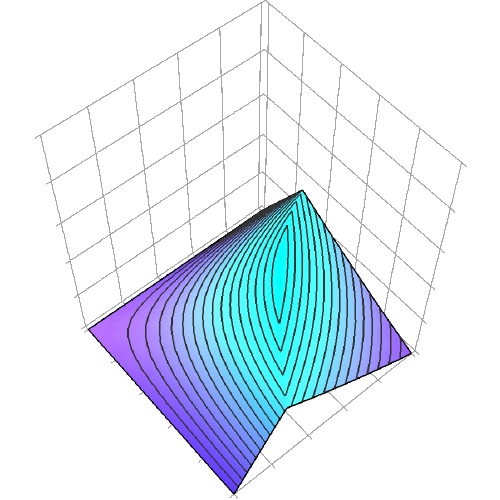}
    \includegraphics[width=\textwidth]{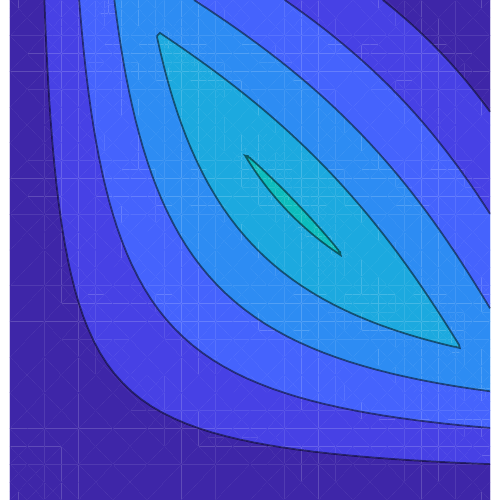}
    \caption{$\psi (\vx) = g (\vx) - | \varphi^h (\vx) |$}
    \label{fig:bimat-psi}
  \end{subfigure}
  \caption{Construction of the enrichment function, $\psi (\vx)$, for the
           material interface problem on a quadratic finite element mesh ($p =
           2$). The plots are on the element domain $\Omega_e = (-0.5,
           -0.25)^2$, with $\varphi (\vx) = || \vx || - 0.5$.}
  \label{fig:bimat-enrichment}
\end{figure}

To model material interfaces using the X-FEM, the normal strain must admit a
jump that satisfies the Hadamard compatibility condition at the interface
between two dissimilar isotropic materials~\cite{Sukumar:2001:MHI}.  Two
additional desirable features of the material interface enrichment function
$\psi (\vx)$ are:
\begin{enumerate}
  \item linear, ridge-like variation near the material interface, and
  \item zero-value outside enriched elements to avoid the effects of blending
  elements~\cite{Chessa:2003:OCB}.
\end{enumerate}
If the enrichment function is not affine, the space of polynomial functions that
can be reproduced is affected, which ultimately degrades the convergence rate of
the method~\cite{Chin:2019:MCI}.  The modified abs-enrichment introduced by
Mo\"{e}s et al.~\cite{Moes:2003:ACA} reproduces both these features for linear
elements, and a modification to this enrichment function to retain this behavior
for higher-order elements with curved interfaces is discussed in Chin and
Sukumar~\cite{Chin:2019:MCI}.  \revtwo{The key features of the enriched element
are that it enables independent $p$-th order polynomials to be reproduced on
both sides of the material interface and that it enforces $C^0$ continuity at
the material interface.}  To define the enrichment function, we first identify
two sets of nodes. Over the element tessellation $\mathcal{T}$, let
$\set{I}_{\text{enr}}$ be the set consisting of nodes that are enriched, i.e.,
the nodal basis function support of these nodes intersects $\Gamma_\text{mat}$
(nodes marked by circles in~\fref{fig:xfem-composites}). We define
$\set{I}_{\text{zero}} \subset \set{I}_{\text{enr}}$ as the set of enriched
nodes whose basis of support also includes elements that do not intersect
$\Gamma_\text{mat}$ (see \fref{fig:set-izero}).  The material interface
enrichment function is set to zero at these nodes to avoid blending effects in
adjacent elements.  For a point $\vx \in \Omega_e$, we define the function 
\begin{equation}\label{eq:bimat-g}
  g^h (\vx) = \sum_{i \in \set{I}_e} N_i (\vx) g_i ,
\end{equation}
where $g_i$ are nodal values which are set by a two-step process.
\begin{enumerate}
  \item For nodes $ i \in \set{I}_{\text{zero}}$, set $g_i = | \varphi_i |$,
    where $\varphi_i := \varphi (\vx_i)$ are nodal values of the signed
    distance function to $\Gamma_\text{mat}$, and for all other nodes set $g_i =
    0$.
  \item Then, for nodes not at the vertices of the element and in
    $\set{I}_{\text{enr}} \backslash \set{I}_{\text{zero}}$, set $g_i =
    \displaystyle\sum_{j = 1}^4 N^{(p=1)}_j (\vx_i) g_j$, where $N^{(p=1)}_j
    (\vx)$ are the bilinear finite element shape functions and $g_j$ are the
    element vertex values of $g$ that are set in step 1.
\end{enumerate}
Using \eqref{eq:bimat-g}, we define our material interface enrichment function
as
\begin{equation}\label{eq:bimat-enrichment}
  \psi (\vx) = g^h(\vx) - | \varphi^h (\vx) | 
    \quad \forall \vx \in \Omega_e ,
\end{equation}
where $\varphi^h (\vx) = \displaystyle\sum_{i \in \set{I}_e} N_i (\vx)
\varphi_i$ is the nodal interpolant of the signed distance function.
In~\eqref{eq:bimat-enrichment}, the second term introduces the ridge with
discontinuous derivative at $\Gamma_\text{mat}$, whereas the first term shifts
the second term to be zero-valued on the appropriate boundaries and approximates
a smoothly interpolated bilinear function elsewhere.  In
\fref{fig:bimat-enrichment}, the enrichment function for a circular inclusion
over quadratic elements is plotted.

\subsection{Discrete equations}\label{ssec:discrete-equations}
For \revtwo{elements cut by $\Gamma_\text{hole}$}, we
choose~\eqref{eq:uh-hole} to form the trial displacement field $\vm{u} (\vx)$
and the test field $\vm{v} (\vx)$ in~\eqref{eq:weak-form-final}.  Then,
considering the arbitrariness of nodal variations, we obtain the element level
quantities:
\begin{equation}\label{eq:hole-element-quantities}
  \vm{K}_e = \int_{\Omega_e \cap \Omega\backslash\Omega_\text{hole}} 
    \big( \vm{B}_u (\vx) \big)^* \vm{D} (\vx) \vm{B}_u (\vx) \, d \vx,  \quad 
  \vm{M}_e = \int_{\Omega_e \cap \Omega\backslash\Omega_\text{hole}} 
    \rho (\vx) \big( \vm{N}_u (\vx) \big)^* \vm{N}_u (\vx) \, d \vx ,
\end{equation}
where $\vm{D} (\vx)$ is the piecewise constant two-dimensional linear elastic
constitutive matrix and the superscript $(\cdot)^*$ denotes the Hermitian of the
matrix.  In addition, $\vm{N}_u (\vx)$ is the standard element shape function
vector and
\begin{equation}\label{eq:b-matrix}
  \vm{B}_\alpha (\vx) = \begin{bmatrix}
    \vm{B}_\alpha^1 (\vx) & \vm{B}_\alpha^2 (\vx) & \dotsc 
      & \vm{B}_\alpha^{|\set{L}|} (\vx) 
  \end{bmatrix},
\end{equation}
where $\alpha = u$, $\set{L} = \set{I}_e$ and
\begin{equation}
  \vm{B}_u^\ell (\vx) = \begin{bmatrix}
    N_{\ell,x} ( \vx ) & 0 \\
    0                  & N_{\ell,y} ( \vx ) \\
    N_{\ell,y} ( \vx ) & N_{\ell,x} ( \vx )
  \end{bmatrix} .
\end{equation}
Note the Hermitian of element matrices $\vm{B}_u (\vx)$ and $\vm{N}_u (\vx)$ is
required on account of the Bloch-periodic boundary conditions
in~\eqref{eq:periodic-dirichlet}.

\revtwo{For elements cut by $\Gamma_\text{mat}$},
choosing~\eqref{eq:uh-bimat} to represent $\vm{u} (\vx)$ and $\vm{v} (\vx)$
in~\eqref{eq:weak-form-final}, and using a standard Galerkin procedure, we
obtain the following element quantities:
\begin{equation}\label{eq:bimat-element-quantities}
    \begin{gathered}
    \vm{K}_e = \bigintsss_{\Omega_e} \begin{bmatrix}
      \big( \vm{B}_u (\vx) \big)^* \vm{D} (\vx) \vm{B}_u (\vx) & 
        \big( \vm{B}_u (\vx) \big)^* \vm{D} (\vx) \vm{B}_a (\vx) \\
      \big( \vm{B}_a (\vx) \big)^* \vm{D} (\vx) \vm{B}_u (\vx) & 
        \big( \vm{B}_a (\vx) \big)^* \vm{D} (\vx) \vm{B}_a (\vx) \\
    \end{bmatrix} \, d \vx , \\
    \vm{M}_e = \bigintsss_{\Omega_e} \rho (\vx) \begin{bmatrix}
      \big( \vm{N}_u (\vx) \big)^* \vm{N}_u (\vx) & 
        \big( \vm{N}_u (\vx) \big)^* \vm{N}_a (\vx) \\
      \big( \vm{N}_a (\vx) \big)^* \vm{N}_u (\vx) & 
        \big( \vm{N}_a (\vx) \big)^* \vm{N}_a (\vx) \\
    \end{bmatrix} \, d \vx ,
  \end{gathered}
\end{equation}
where $\vm{B}_\alpha (\vx)$ for $\alpha = u, a$ is defined in
\eqref{eq:b-matrix} and $\vm{N}_a (\vx)$ is the element shape function vector
for nodes in $\set{J}_e$.  
Referring to~\eqref{eq:b-matrix}, for $\alpha = a$ and $\set{L} = \set{J}_e$,
we define
\begin{equation}
  \vm{B}_a^\ell (\vx) = \begin{bmatrix}
    \left( N_\ell ( \vx ) \psi (\vx) \right)_{,x} & 0 \\
    0 & \left( N_\ell ( \vx ) \psi (\vx) \right)_{,y} \\
    \left( N_\ell ( \vx ) \psi (\vx) \right)_{,y} & 
      \left( N_\ell ( \vx ) \psi (\vx) \right)_{,x}
  \end{bmatrix} .
\end{equation}

\revtwo{For the remaining elements that are not in $\Omega_\text{hole}$, no
enrichment is required.  In these elements, the standard finite element
displacement field in~\eqref{eq:uh} is used to approximate $\vm{u}(\vx)$ and
$\vm{v}(\vx)$ in~\eqref{eq:weak-form-final}.  The resulting element level
quantities are:
\begin{equation}\label{eq:element-quantities}
  \vm{K}_e = \int_{\Omega_e} 
    \big( \vm{B}_u (\vx) \big)^* \vm{D} (\vx) \vm{B}_u (\vx) \, d \vx,  \quad 
  \vm{M}_e = \int_{\Omega_e} 
    \rho (\vx) \big( \vm{N}_u (\vx) \big)^* \vm{N}_u (\vx) \, d \vx .
\end{equation}}

Assembling element level quantities using the finite element assembly procedure,
the generalized eigenproblem, $\vm{K d} = \lambda \vm{M d}$, is obtained.
Element-level integration over elements intersected by holes and material
interfaces is conducted using the HNI method, which requires homogeneous
integrands.  Details of integration using the HNI method are described in
\sref{ssec:numerical-integration}.  \revtwo{Integration
of~\eqref{eq:element-quantities} is accomplished using a tensor-product
Gauss-Legendre quadrature rule.}

To generate a band structure diagram, the Bloch-periodic weak form is solved for
many wave vectors, $\vm{k}$, within the irreducible Brillouin zone of the unit
cell.  Note the vector of nodal displacements, $\vm{d}$, that satisfies
Bloch-periodic boundary conditions given in \eqref{eq:fe-space} is related to
the unconstrained vector of nodal displacements, $\hat{\vm{d}}$,
via~\cite{Alberdi:2018:AIA}
\begin{subequations}\label{eq:bloch-bc-matrix}
\begin{equation}
  \hat{\vm{d}} := \vm{R} \vm{d},
\end{equation}
where
\begin{equation}
  \vm{R}_{IJ} = \begin{cases}
    \vm{I} \exp(i \vm{k} \cdot \vm{h}_i) & \text{if } \vm{d}_I \in \mathbb{D}_i \\
    \vm{I} \delta_{IJ} & \text{otherwise}
  \end{cases}
\end{equation}
\end{subequations}
is a $2 \times 2$ block in the constraint matrix $\vm{R}$
associated with unconstrained degree of freedom $I$ and constrained degree of
freedom $J$, and $\mathbb{D}_i$ is the set of nodes located on $\Gamma_i$, for
$i = 1, 2$ (see \fref{fig:unit-cell-bvp}).  In \eqref{eq:bloch-bc-matrix},
$\vm{R}$ is a $\bigl[ndof \times 2\bigr] \times \bigl[ (ndof - |\mathbb{D}_1| -
|\mathbb{D}_2| + 1) \times 2 \bigr]$ matrix, where $ndof$ is the number of
unconstrained DOFs.  The constraint matrix allows the finite element system of 
equations to be equivalently expressed as
\begin{equation}
  \vm{R}^* \hat{\vm{K}} \vm{R} \vm{d} = 
    \lambda \vm{R}^* \hat{\vm{M}} \vm{R} \vm{d} ,
\end{equation}
where $\hat{\vm{K}}$ and $\hat{\vm{M}}$ are the global stiffness and mass
matrices, respectively, with no Bloch-periodic boundary conditions.  Thus,
$\hat{\vm{K}}$ and $\hat{\vm{M}}$ can be formed once for an extended finite
element system of equations, then the matrix $\vm{R}$ can be formed to generate
the constrained equations for each $\vm{k}$ in the band structure diagram. While
not explored in this paper, secondary acceleration techniques such as
RBME~\cite{hussein2009reduced} and Bloch mode
synthesis~\cite{krattiger2014bloch} can also be applied to further reduce the
computational burden of computing frequencies at each $k$-point.

For one-dimensional material interface problems, the trial eigenfunction in
spectral finite elements is:
\begin{align}
u^h(x) &= \sum_{j \in \mathbb{J}} \phi_j(x) u_j. 
\end{align}
where $\mathbb{J}$ is the index set of all nodes in $\Omega$ and $\phi_j (x)$ is
the spectral finite element basis function associated with node $j$.  On
choosing $v^h = \phi_i$ as the test eigenfunction, we obtain the following
generalized eigenproblem:
\begin{subequations}
\begin{align}
\vm{K}\vm{d} &= \lambda \vm{M} \vm{d}, \\
\vm{K}_{ij} = \int_0^a \phi_i^{*\prime} E \phi_j^\prime \, dx &,
\quad
\vm{M}_{ij} = \int_0^a \phi_i^* \rho \phi_j \, dx ,
\end{align}
\end{subequations}
where $\phi_i^*$ denotes the complex conjugate of $\phi_i$.

\section{Representation of curved geometries}\label{sec:curved-geometries}

Two methods of defining curved material and void boundaries are considered in
this paper.  The first is through an implicit representation of the boundary
using level set functions.  The second is through quadratic rational \bezier{}
curves, which provide a parametric description of conic boundaries.
\revtwo{Explicit, parametric descriptions of the material interface and boundary
locations are required to invoke the HNI method.}  The remainder of this section
provides relevant details required to solve a problem using the X-FEM with these
representations of curved boundaries.
\subsection{Level set representation}\label{ssec:level-set}
In the level set method, isocontours of a $(d + 1)$-dimensional function
$\varphi (\vx)$ are used to track the location of geometric interfaces within a
$d$-dimensional domain.  We select the level curve $\varphi (\vx) = 0$ as the
interface location.  Further, $\varphi (\vx) < 0$ and $\varphi (\vx) > 0$ denote
regions that lie inside the closed interface and outside the closed interface,
respectively.  An example of a level set function is the signed distance
function, which is used in the material interface enrichment developed in
\sref{ssec:modeling-interfaces}.  The level set method simplifies determination
of nodal locations with respect to the interface, which simplifies identifying
elements that require enrichment in the X-FEM.  Consistent with the material
interface enrichment in~\eqref{eq:bimat-enrichment}, we approximate the level
set function at the element level with its nodal interpolant:
\begin{equation}
  \varphi^h (\vx) = \sum_{i \in \set{I}_e} N_i (\vx) \varphi_i .
\end{equation}

To integrate the weak form integrals in \eqref{eq:weak-form-final} using the HNI
method introduced in \sref{ssec:numerical-integration}, an explicit
representation of the interface is required.  We employ cubic Hermite functions
for this purpose.  For the parameter $t \in [0,1]$, a cubic Hermite function is
\begin{equation}\label{eq:hermite}
  \vm{c} (t, \vm{m}) = (2t^3 - 3t^2 + 1) \vm{p}_0 + (t^3 - 2 t^2 + t) m_0 \vm{t}_0 
  + (-2t^3 + 3t^2) \vm{p}_1 + (t^3 - t^2) m_1 \vm{t}_1 ,
\end{equation}
where $\vm{m} = \left\{ m_0, m_1 \right\}$, $\vm{p}_0$ and $\vm{p}_1$ are the
endpoints of the curve, and $m_0 \vm{t}_0 $ and $m_1 \vm{t}_1$ are the end
tangents of the curve at $t = 0$ and $t = 1$, respectively.  The endpoints
($\vm{p}_0$ and $\vm{p}_1$) and tangent vectors ($\vm{t}_0$ and $\vm{t}_1$) of
the curve are immediately identified by computing $\varphi^h (\vx)$ and
$-\bigl(\nabla \varphi^h (\vx)\bigr)^\perp$, where $\bigl(\nabla \varphi^h
(\vx)\bigr)^\perp := \Bigl[ \bigl( \partial \varphi^h (\vx) \bigr) / (\partial
y) \ \ - \bigl(\partial \varphi^h (\vx) \bigr) / (\partial x) \Bigr]^T$ is
$\nabla \varphi^h (\vx)$ rotated through $-\pi / 2$ radians, at the endpoints.
The magnitude of the tangent vector, $m_0$ and $m_1$, remain to be determined.

The values $m_0$ and $m_1$ should be selected to approximate the zero isocontour
with minimal error.  Exact Hermite reconstruction recovers (weakly)
\begin{equation}\label{eq:exact-hermite-cond}
  F(\vm{m}) := \int_0^1 
    \left[ \varphi^h \bigl(\vm{c}(t, \vm{m})\bigr) \right]^2 \, dt = 0 .
\end{equation}
Following~\cite{Chin:2019:MCI}, we choose optimal values of $m_0$ and $m_1$ by
solving
\begin{equation}\label{eq:objective-fn}
  \vm{m}^* = \argmin_{\vm{m} \in \Re^2} \ F(\vm{m}) .
\end{equation}
This numerical optimization problem can be solved using techniques such as
Newton's method or the Broyden-Fletcher-Goldfarb-Shanno (BFGS) algorithm.  The
BFGS algorithm does not require computing second derivatives of $\varphi^h
(\vx)$, enhancing its appeal in the context of finite elements since second
derivatives of shape functions are typically not computed.

If $F (\vm{m}^*) > 0$, then Hermite reconstruction of the isocontour is not
exact.  Total error over the path is estimated by~\cite{Chin:2019:MCI}
\begin{equation}\label{eq:ls-approx-error}
  e (\vm{m}^*) = \sqrt{\int_0^1 \left(\frac{\varphi^h \bigl(\vm{c} (t, \vm{m}^*)\bigr)}
    {\big|\big| \nabla \varphi^h \bigl(\vm{c} (t, \vm{m}^*)\bigr) \big|\big|} 
    \right)^2 
    \bigg|\bigg| \frac{\partial \vm{c} (t, \vm{m}^*)}
    {\partial t} \bigg|\bigg| \, dt } .
\end{equation}
The value $e (\vm{m}^*)$ is compared to a user-defined tolerance, $\epsilon$.
If $e (\vm{m}^*)$ exceeds $\epsilon$, the Hermite curve can be recursively
bisected until $e (\vm{m}^*) < \epsilon$ in all curve segments.  This
methodology provides a measure of adaptive refinement, optimizing the number and
location of Hermite curves needed to accurately trace the implicit curve
$\varphi^h(\vx) = 0$.
\subsection{Rational \bezier{} representation}
Rational quadratic \bezier{} curves are capable of exactly representing conic
sections such as circles and ellipses.  Given \bezier{} control points
$\vm{p}_0$, $\vm{p}_1$, and $\vm{p}_2$ and a coordinate $\vx$, the \bezier{}
curve can be transformed to a level set function by considering its barycentric
representation.  Note this level set function is not equal to the signed
distance function and, therefore, is not a good candidate for use with the
material interface enrichment in~\eqref{eq:bimat-enrichment}.  As described in
Farin~\cite{Farin:1996:CSC}, the linear system of equations,
\begin{align*}
  \tau_0 \vm{p}_0 + \tau_1 \vm{p}_1 + \tau_2 \vm{p}_2 & = \vx , \\
  \tau_0 + \tau_1 + \tau_2 & = 1 ,
\end{align*}
are solved for parameters $\tau_0$, $\tau_1$, and $\tau_2$.  Then, the value of
the implicit function is given by
\begin{equation*}
  \varphi (\vx) = \tau_1^2 - 4 \frac{\tau_0 \tau_2 w_1^2}{w_0 w_2},
\end{equation*}
where $w_0$, $w_1$, and $w_2$ are weights associated with the control points
$\vm{p}_0$, $\vm{p}_1$, and $\vm{p}_2$, respectively.

\section{Numerical implementation of the X-FEM}\label{sec:xfem-implementation}
\subsection{Numerical integration}\label{ssec:numerical-integration}
Discontinuous integrands in \eqref{eq:hole-element-quantities} and
\eqref{eq:bimat-element-quantities} are ill-suited for integration using a
tensor-product Gauss rule.  To accurately and efficiently integrate the
discontinuous weak form integrals, we employ the HNI method. The HNI method
traces its origins to Lasserre~\cite{Lasserre:1998:ICP}, who used Euler's
homogeneous function theorem to simplify integration over a $d$-dimensional
convex polytope to integration over the $(d-1)$-dimensional faces of the
polytope.  Chin et al.~\cite{Chin:2015:NIH} extended Lassere's approach to
nonconvex regions.  This section gives a broad overview of the method for two
dimensional domains of integration.  We refer the reader to Chin and
Sukumar~\cite{Chin:2019:MCI} for a thorough examination of implementation
details.

Let $f(\vx)$ be a continuously differentiable, positively homogeneous function
of degree $q$.  Our objective is to integrate $f(\vx)$ over a domain $A \subset
\Omega_e$, i.e.,
\begin{equation}
  I = \int_A f(\vx) \, d \vx .
\end{equation}
The weak form integrands in \sref{ssec:discrete-equations} are polynomial
functions, which can be decomposed into a sum of homogeneous functions. Further,
when the domain of integration in enriched elements is split where the
integrands are discontinuous, we recover subregions whose boundaries are
composed of a combination of affine and curved edges.  When $f(\vx)$ is a homogeneous polynomial, Euler's homogeneous function theorem states that
\begin{equation}\label{eq:euler}
  q f(\vx) = \nabla f(\vx) \cdot \vx \quad \forall \vx \in \Re^2.
\end{equation}
For a vector field $\vm{X} := \vm{X} (\vx)$, Stokes's theorem can be stated as:
\begin{equation}\label{eq:stokes}
  \int_A ( \nabla \cdot \vm{X} ) f (\vx) \, d \vx + 
  \int_A \nabla f (\vx) \cdot \vm{X} \, d \vx 
  = \int_{\partial A} ( \vm{X} \cdot \vm{n} ) f(\vx) \, d s ,
\end{equation}
where $\partial A$ is the boundary of $A$ and $d s$ is differential arc length
on $\partial A$. Invoking~\eqref{eq:euler} on homogeneous function $f (\vx)$
and choosing $\vm{X}$ as the position vector $\vx$, \eqref{eq:stokes} yields
\begin{equation} \label{eq:finaleq}
  \int_A f (\vx) \, d \vx = \frac{1}{2 + q} \sum_{i=1}^m 
  \int_{F_i} (\vx \cdot \vm{n}_i) f (\vx) \, d s ,
\end{equation}
where $\partial A := \overline{F_1 \cup F_2 \cup \ldots \cup F_m}$.
Equation~(\ref{eq:finaleq}) relates integration of a positively homogeneous
function $f(\vm{x})$ over a domain in $\Re^2$ to integration over the domain's
one-dimensional boundary.  We assume that $\partial A = \overline{\partial A_A
\cup \partial A_C}$, where $\partial A_A$ is a collection of line segments and
$\partial A_C$ is a set consisting of parametric curves.

On $\partial A_A$, each line segment is the subset of a line given by the
equation $\vm{a} \cdot \vm{x} = b$.  The unit normal of the line is the
normalized gradient of this equation, i.e., $\vm{n} = \vm{a} / || \vm{a} ||$.
The sign of $\vm{a}$ and $b$ are selected such that $\vm{n}$ points outward
from $A$.  Substituting into \eqref{eq:finaleq}, we obtain
\begin{equation}\label{eq:affine}
  \int_{F_i \subset \partial A_A} (\vx \cdot \vm{n}_i) f (\vx) \, d s 
  = \frac{b_i}{|| \vm{a}_i ||} \int_{F_i \subset \partial A_A} 
    f (\vx) \, d s .
\end{equation}
On $\partial A_C$, we consider parametric curves of the form $\vm{c} (t)$, where
$t \in [0, 1]$ is a parameter such that $\vm{c} (0)$ and $\vm{c} (1)$ give the
endpoints of $F_i \subset \partial A_C$.  The curve parameter is chosen such
that the boundary is traversed counterclockwise as $t$ increases.  We note
$\vm{c} (t)$ encompasses many types of curves, including the rational \bezier{}
curves and Hermite spline curves discussed in \sref{sec:curved-geometries}.
Substituting into \eqref{eq:finaleq}, we obtain
\begin{equation}\label{eq:parametric1}
  \int_{F_i \subset \partial A_C} (\vx \cdot \vm{n}_i) f(\vx) \, d s 
  = \int_0^1 \left( \vm{c}_i (t) \cdot \vm{n}_i \right) f \left( \vm{c}_i (t)
  \right) || \vm{c}'_i (t) || \, dt ,
\end{equation}
where $\vm{c}'_i (t)$ is the derivative with respect to the parameter $t$, i.e.,
the hodograph of the curve.  The normal of $\vm{c} (t)$ is the unit tangent
vector rotated counterclockwise through $-\pi / 2$ radians.  Substituting this
into \eqref{eq:parametric1} recovers
\begin{equation}\label{eq:parametric}
  \int_{F_i \subset \partial A_C} (\vx \cdot \vm{n}_i) f (\vx) \, d s
  = \int_0^1 \left( \vm{c}_i (t) \cdot \vm{c}^{\prime\perp} (t)
  \right) f \left( \vm{c}_i (t) \right) dt .
\end{equation}
With \eqref{eq:affine} and \eqref{eq:parametric}, integration over $A$ reduces
to one-dimensional integrals over $\partial A$.  These integrals are computed
using Gauss quadrature.  Note if $\vm{c} (t)$ and $f (\vx)$ are polynomial,
integration is exact with an appropriate quadrature rule.

We use \eqref{eq:finaleq}, \eqref{eq:affine}, and \eqref{eq:parametric} to
develop a cubature rule to apply to the polynomial weak form integrals
in~\sref{ssec:discrete-equations}.  To simplify implementation, we develop a
single cubature rule for each element designed to integrate the highest degree
polynomial in the integrand.  This integration rule is therefore valid on all
homogeneous terms of the integrand.  On affine edge $F_j$, cubature points are
Gauss points mapped to $F_j$ and cubature weights are selected as $\tilde{w}_i
b_j / || \vm{a}_j ||$ for $i = 1, \ldots, m_j$, where $\tilde{w}_i$ are Gauss
weights of an $m_j$-point rule scaled by the length of $F_j$. On curved edge
$F_k$, cubature points are $\vm{x}_i = \vm{c}_k (t_i)$, where $t_i$ ($i = 1,
\ldots, m_k$) are Gauss points of a $m_k$-point rule mapped to the interval
$[0,1]$ and cubature weights are $w_i = \left( \vm{c}_k (t_i) \cdot
\vm{c}_k^{\prime\perp} (t_i) \right) \tilde{w}_i$ for $i = 1, \ldots, m_k$,
where $\tilde{w}_i$ are Gauss weights of an $m_k$-point rule scaled to the
interval $[0,1]$.

We decompose a polynomial function, $h(\vx)$, into $nh$ $q_j$-homogeneous
polynomial functions: $f_j (\vx)$ for $j = 1, \dotsc, nh$, such that
\begin{equation}
  h(\vx) = \sum_{j=1}^{nh} f_j(\vx) .
\end{equation}
Combining the contributions over each boundary edge, the cubature rule is then
\begin{equation}\label{eq:hni-quad-rule}
  I = \int_{\Omega_e\backslash\Omega_\text{hole}} h (\vx) \, d \vx \approx 
  \sum_{i = 1}^{nq} \sum_{j = 1}^{nh} \frac{1}{2 + q_j} f_j (\vx_i) w_i ,
\end{equation}
where $nq$ is the total number of cubature points, $\left\{ \vx_i \right\}_{i =
1}^{nq}$ are the cubature points, and $\left\{ w_i \right\}_{i = 1}^{nq}$ are
the cubature weights.

Applying \eqref{eq:hni-quad-rule} to \eqref{eq:hole-element-quantities} using a
$p$-th order finite element, we obtain
\begin{subequations}
\begin{align}
  \vm{K}_e & \approx \sum_{i=1}^{nq} \left( 
    \frac{\vm{K}_e^{[0]}}{2} + 
    \frac{\vm{K}_e^{[1]} (\vx_i)}{3} +
    \cdots +
    \frac{\vm{K}_e^{[4p-2]} (\vx_i)}{4p} \right) w_i ,
  \intertext{and}
  \vm{M}_e & \approx \sum_{i=1}^{nq} \left( 
    \frac{\vm{M}_e^{[0]}}{2} + 
    \frac{\vm{M}_e^{[1]} (\vx_i)}{3} +
    \cdots +
    \frac{\vm{M}_e^{[4p]} (\vx_i)}{4p + 2} \right) w_i ,
\end{align}
\end{subequations}
where $\vm{K}_e^{[q]}$ and $\vm{M}_e^{[q]}$ denote the degree-$q$ term in the
homogeneous expansions of the integrands in \eqref{eq:hole-element-quantities}.
The weak form integrals for an element with interface enrichment can be
similarly decomposed to a form amenable to integration using
\eqref{eq:hni-quad-rule}.
\subsection{Interface stabilization}\label{ssec:interface-stabilization}
The extended finite element approximation for holes and material inclusions can
induce high condition number in the global finite element stiffness matrix,
which can lead to reduced accuracy when solving the algebraic system of linear
equations~\cite{Babuska:2017:SSG}.  The reason for poor matrix conditioning is
distinct in the hole problem versus the material inclusion problem.  For voids,
elements where $\Omega_e \cap \Omega_\text{hole}$ is very large compared to
$\Omega_e$ cause very small regions of support on some degrees of freedom. These
ultimately result in modes of deformation with much lower stiffness and inertia
when compared to others, resulting in an ill-conditioned system of linear
equations. In the material inclusion problem, conditioning issues arise when
$\Omega_e \cap \Omega_{I}$ or $\Omega_e \cap \Omega_{II}$ is very small compared
to $\Omega_e$ or when the scaling of the enrichment function substantially
differs from the shape functions.  A summary of methods to improve conditioning
in the X-FEM is discussed in de Prenter et al.~\cite{dePrenter:2017:CNA}.

\begin{figure}
  \centering
  \begin{subfigure}{3.2in}
    \includegraphics[scale=0.969]{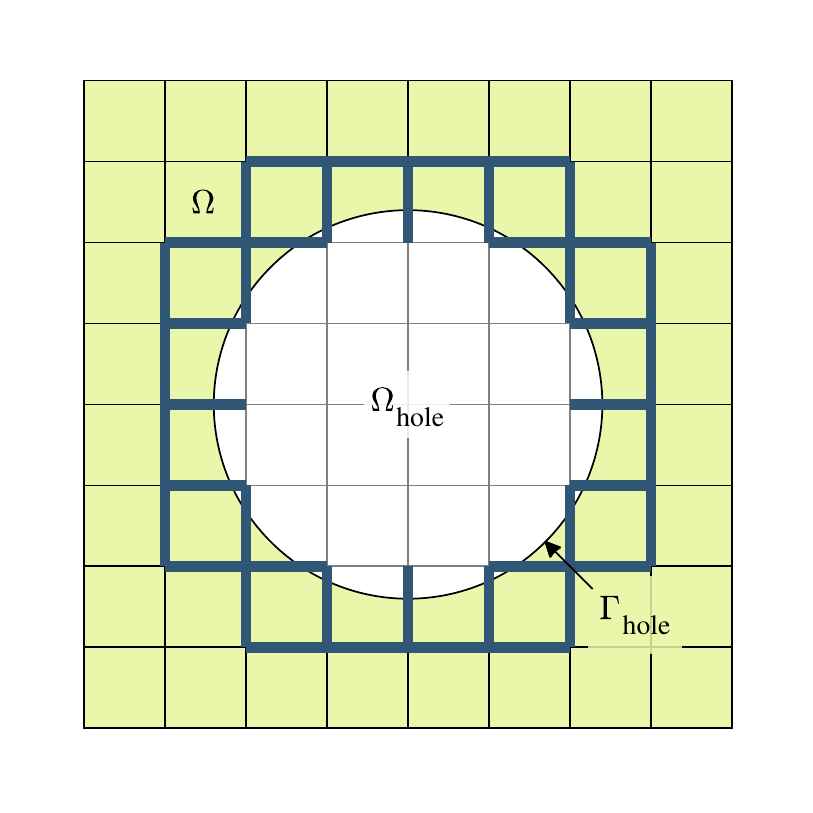}
    \caption{}\label{fig:ghost-penalty-edges-1}
  \end{subfigure}
  \begin{subfigure}{3.2in}
    \includegraphics[scale=0.969]{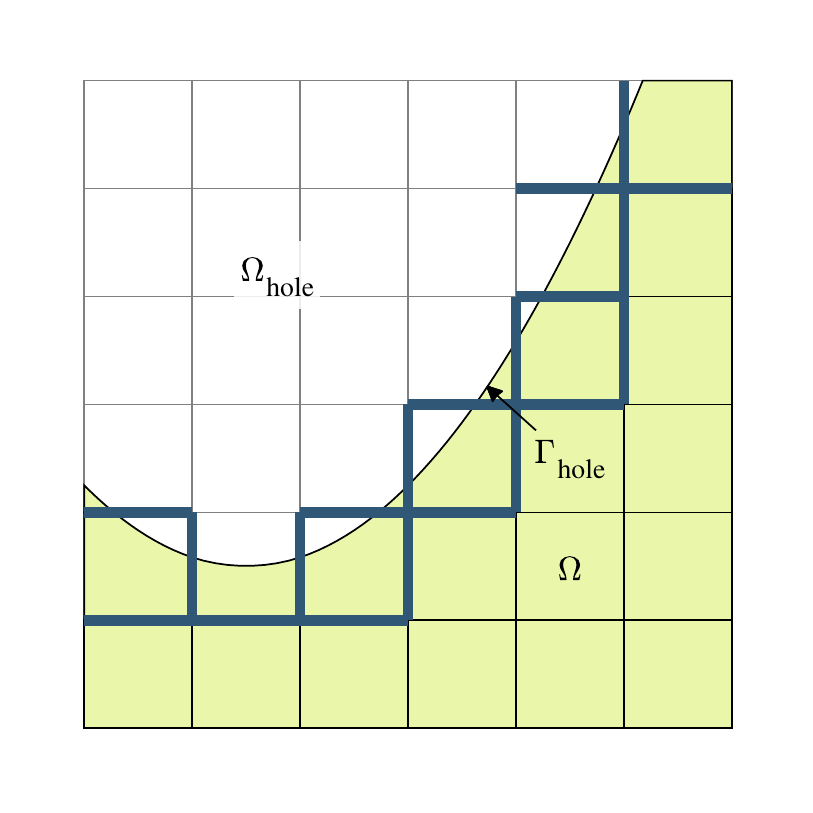}
    \caption{}\label{fig:ghost-penalty-edges-2}
  \end{subfigure}
  \caption{Illustrating the set of edges $\mathcal{F}$ on which the ghost
           penalty stabilization term is computed.  Edges in $\mathcal{F}$ are
           bold. The element tessellation, $\mathcal{T}$, is an $8 \times 8$ and
           a $6 \times 6$ mesh of square elements in (a) and (b), respectively.
           }
  \label{fig:ghost-penalty-edges}
\end{figure}

To provide coercivity over the computational domain for the hole problem, we
introduce a \textit{ghost penalty} stabilization
energy~\cite{Burman:2010:GP,Burman:2015:CDG}.  The ghost penalty term is
designed to eliminate the influence of very small cut elements on the condition
number of the stiffness matrix.  The term adds additional stiffness and mass
contributions to DOFs that have some support in $\Omega_\text{hole}$, which
minimally changes the finite element solution. We establish the set of elements
in $\Omega$ and the set of cut elements as
\begin{equation*}
  \mathcal{T}_a = \Bigl\{ \; T_i \; : \; T_i \in \mathcal{T}, \,
    \Omega_i \cap \Omega \neq \emptyset \; \Bigr\}, \quad 
  \mathcal{T}_c  = \Bigl\{ \; T_i \; : \; T_i \in \mathcal{T}_a, \,
    \Omega_i \cap \Omega \neq \Omega_i \; \Bigr\},
\end{equation*}
where $T_i$ is the $i$-th element in $\mathcal{T}$ with domain $\Omega_i$.
Additionally, we define a set of edges
\begin{equation*}
  \mathcal{F} = \Bigl\{ \; F = T_i \cap T_j \, : \, T_i \in 
    \mathcal{T}_a , \, T_j \in \mathcal{T}_c \; \Bigr\}
\end{equation*}
for $i, j = 1, \ldots, M$.  The set of edges $\mathcal{F}$ is demonstrated for
two domains in \fref{fig:ghost-penalty-edges}.  Following Sticko et
al.~\cite{Sticko:2018:HOC}, the supplemental weak form term in
\eqref{eq:weak-form-final} is
\begin{equation}\label{eq:ghost-penalty}
  j( \vm{u}, \vm{v} ) = \sum_{F \in \mathcal{F}} \sum_{k = 1}^p
    \frac{h^{2k+1}}{ \left( 2k + 1 \right) \left( k! \right)^2} \int_F \left[
    \partial^k_{\vm{n}} \vm{u} \right] \left[ \partial^k_{\vm{n}} \vm{v}
    \right] dS ,
\end{equation}
where $\vm{n}$ denotes a unit vector normal to $F$, $p$ is the order of the
finite element approximation, and $h$ is the average characteristic element
length for the two elements connected to $F$.  Also appearing
in~\eqref{eq:weak-form-final} are the constants $\gamma_K := \beta_K
(\tilde{\lambda} + 2 \mu)$ and $\gamma_M := \beta_K \rho$, where
$\tilde{\lambda}$ and $\mu$ are Lam\'{e} parameters for the material, $\rho$ is
the density of the material, and $\beta_K$ and $\beta_M$ are constants that
scale the magnitude of the ghost penalty terms.  The bracket $[ \cdot ]$ is used
to represent the difference of its argument evaluated over the elements attached
to $F$.

\begin{figure}
  \centering
  \begin{subfigure}{3.2in}
    \includegraphics[scale=0.969]{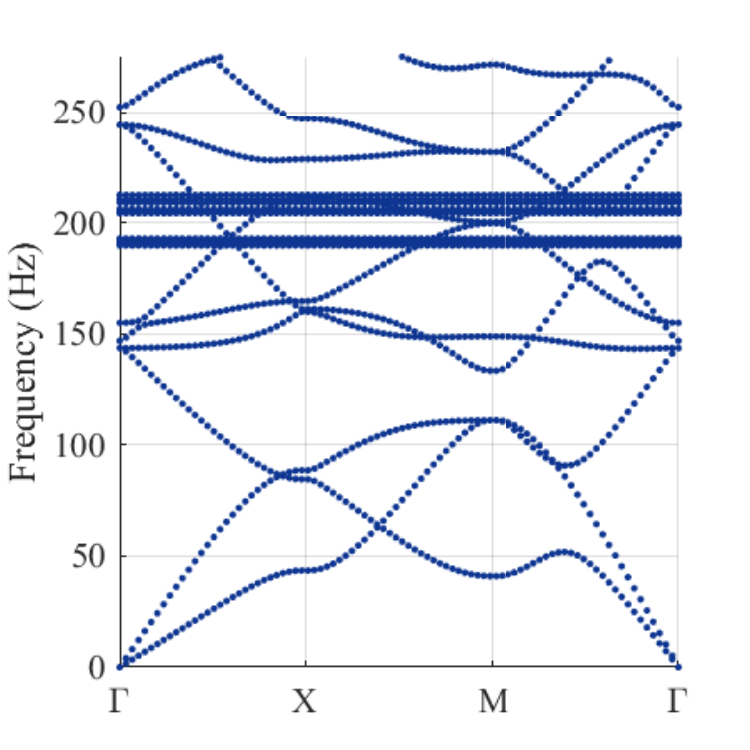}
    \caption{}\label{fig:gp-pollution-polluted}
  \end{subfigure}
  \begin{subfigure}{3.2in}
    \includegraphics[scale=0.969]{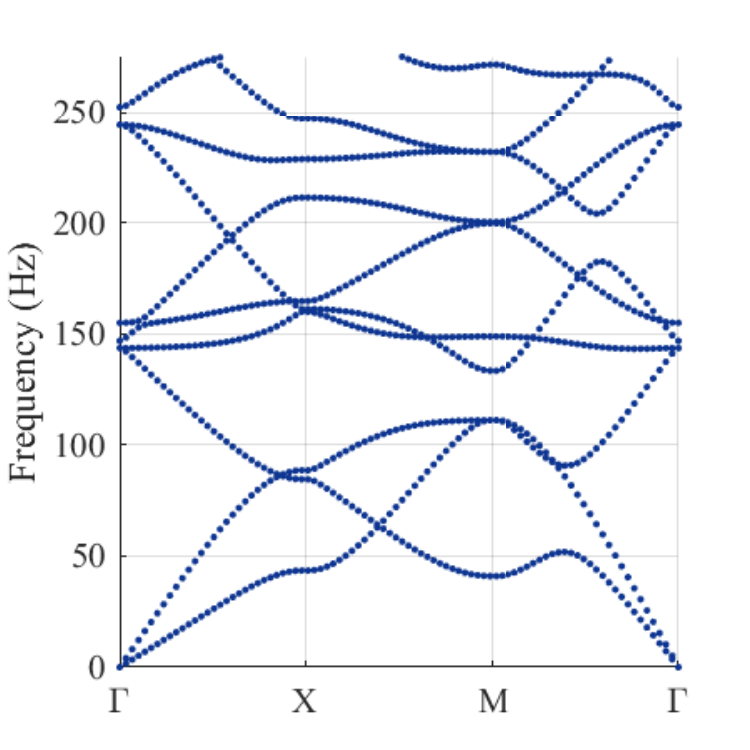}
    \caption{}\label{fig:gp-pollution-fixed}
  \end{subfigure}
  \caption{The effects of ghost penalty modes of deformation on the resulting
    band structure diagram.  (a) Frequencies associated with the ghost penalty
    term appear in the band structure (horizontal lines) and (b) removing these
    modes by increasing the ratio $\beta_K / \beta_M$ by a factor of 4.}
  \label{fig:gp-pollution}
\end{figure}

While the ghost penalty term improves matrix-conditioning, which enables more
accurate calculation of eigenpairs using iterative or direct methods, it also
introduces modes of deformation that can pollute the resulting band structure
diagram. Frequencies of modes of deformation influenced by ghost penalty
stabilization will appear in the band structure diagram if elastic wave speeds
of the ghost penalty modes are less than the wave speeds of the non-ghost
penalty modes of deformation.  If ghost penalty modes appear in the band
structure diagram, the ratio of $\beta_K$ to $\beta_M$ can be increased.  An
example of these modes polluting a band structure is demonstrated in
\fref{fig:gp-pollution-polluted}.  When $\beta_M$ is reduced by a factor of 4,
the polluted modes no longer appear, as \fref{fig:gp-pollution-fixed} reveals.

\section{Numerical results}\label{sec:results}

In this section, several examples are presented that demonstrate the
capabilities of the X-FEM in modeling phononic crystals with voids and material
interfaces and in generating band structure diagrams of these phononic crystals.
We will focus on one-dimensional examples and two-dimensional examples that
contain curved interfaces.  Extended finite element solutions are generated from
a code developed and run in MATLAB 9.9.0.  The function \texttt{eigs()} is
utilized to compute frequencies that satisfy the dispersion relationship for
each periodic domain.  Two-dimensional reference finite element solutions are
generated using the multiphysics package COMSOL.  In all extended finite element
examples with voids, ghost penalty parameters of $\beta_K = 1.0 \times 10^{-8}$
and $\beta_M = 2.5 \times 10^{-9}$ are used.
\subsection{Phononic crystal with circular void and square domain}%
\begin{figure}
  \centering
  \includegraphics[scale=0.969]{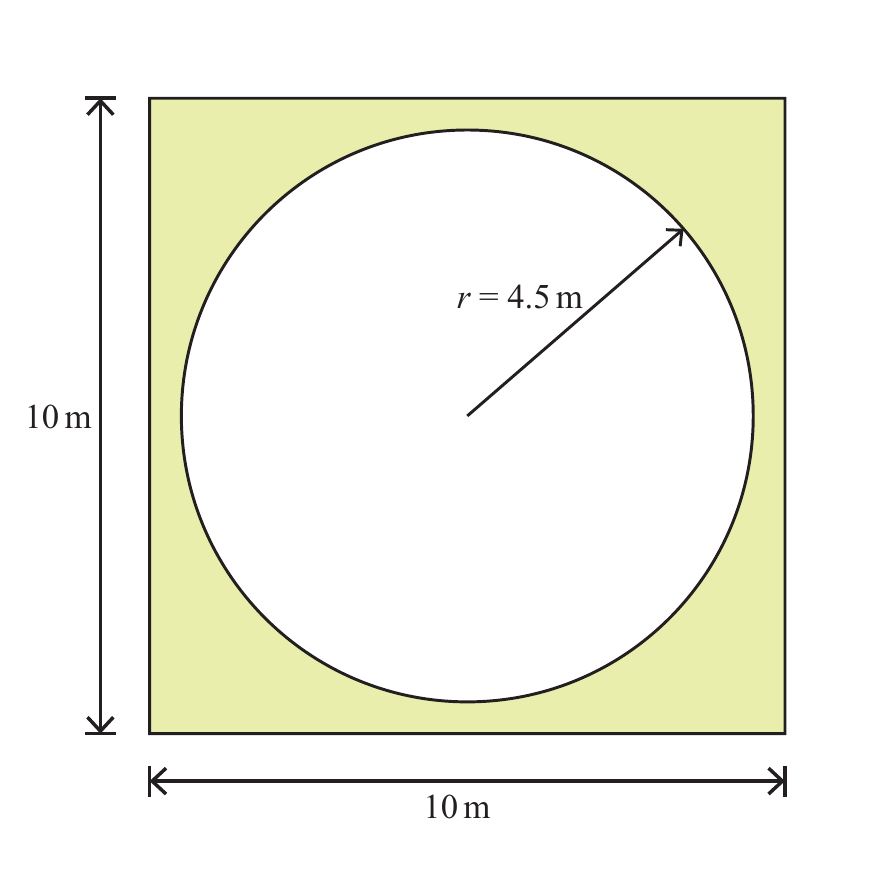}
  \caption{Unit cell geometry of a phononic crystal with a circular void.}
  \label{fig:ex-void-circle-setup}
\end{figure}

\begin{figure}
  \centering
  \includegraphics[scale=0.969]{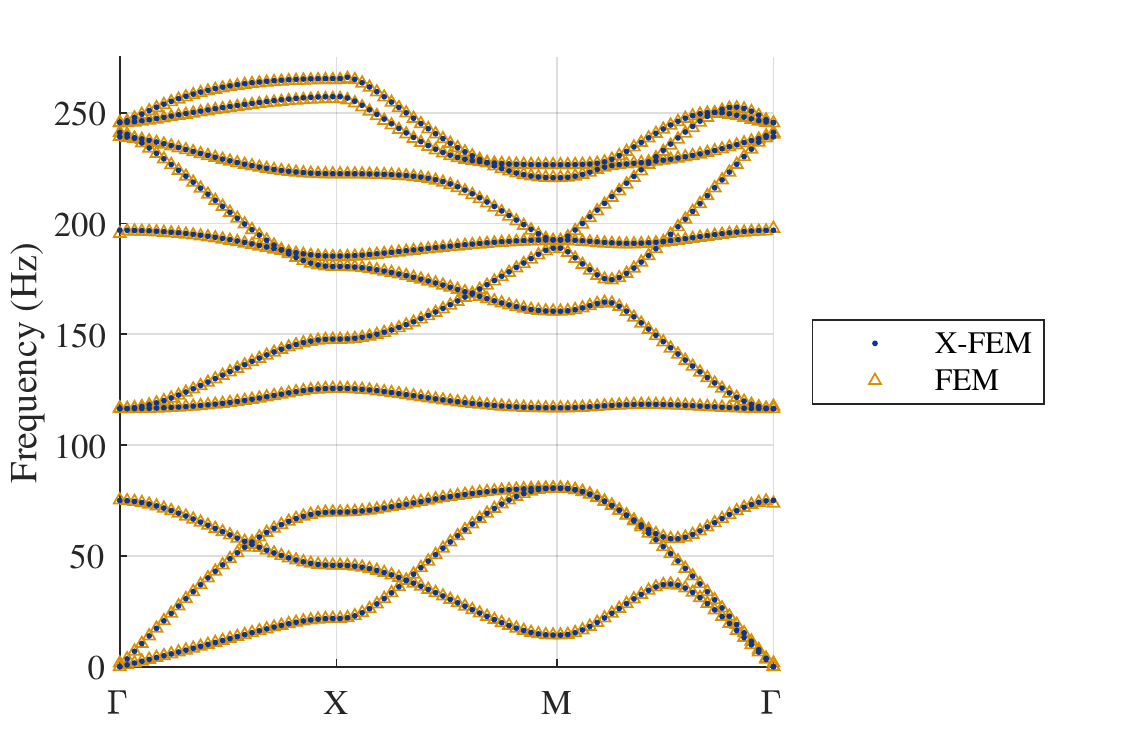}
  \caption{Band structure diagram for a phononic crystal with a circular void.
    Solution using quartic X-FEM (414 DOFs) and reference solution using
    quadratic FEM (227,824 DOFs).}
  \label{fig:ex-void-circle-bsd}
\end{figure}

\label{ssec:ex-hole-circle}
In this example, we develop band structure diagrams for a phononic crystal with
a circular void.  This example has appeared in Wang et al.~\cite{Wang:2011:LBT},
and is repeated here using both the X-FEM and FEM.  Using the X-FEM, the circle
is exactly captured using a quadratic rational B\'{e}zier curve, whereas the FEM
only approximates the shape of the circle.  The square Bloch-periodic unit cell
for this problem is illustrated in \fref{fig:ex-void-circle-setup}.  The
material properties are $E = 20\ \si{\giga\pascal}$, $\nu = 0.25$, and $\rho =
2700\ \si{\kilo\gram\per\meter\cubed}$.  The band structure diagram constructed
using the X-FEM and FEM is presented in \fref{fig:ex-void-circle-bsd}.  The
reference FEM band structure is generated from a finite element mesh with
227,824 DOFs (56,102 quadratic triangular finite elements) while the extended
finite element band structure is generated from 16 quartic square finite
elements (480 DOFs).  The two solutions are observed to be nearly identical.

\begin{figure}
  \centering
  \begin{subfigure}{3in}
    \includegraphics[scale=0.969]{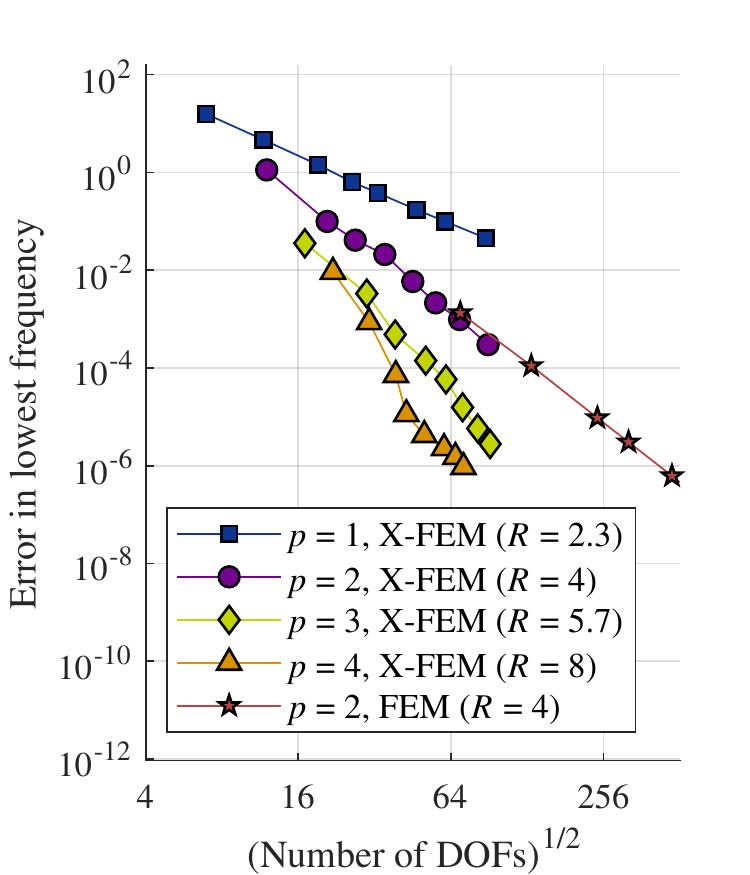}
    \caption{}\label{fig:ex-void-circle-conv-M-1}
  \end{subfigure}
  \begin{subfigure}{3in}
    \includegraphics[scale=0.969]{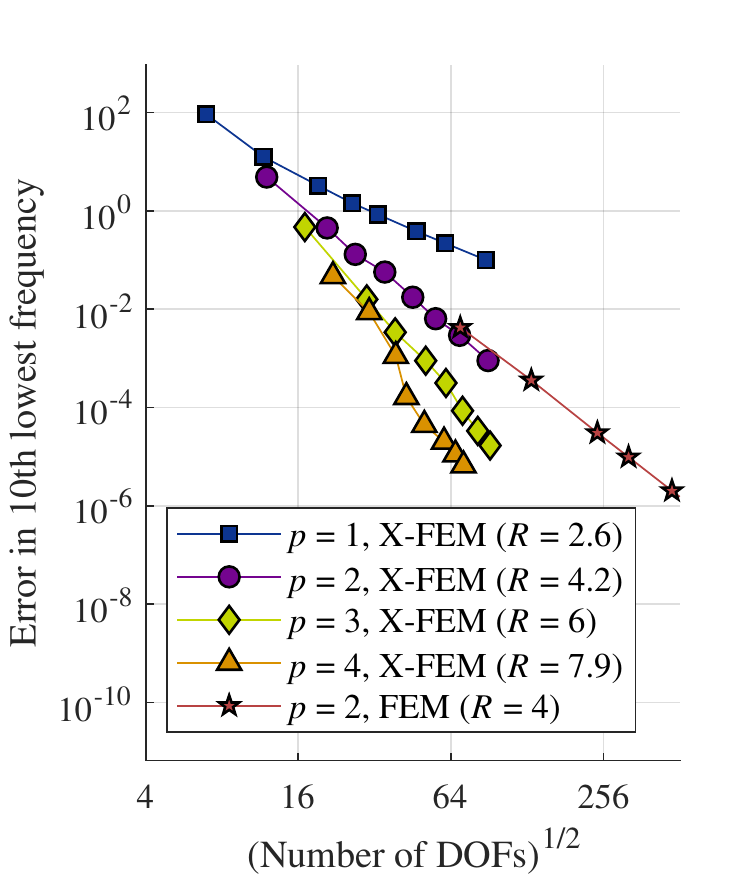}
    \caption{}\label{fig:ex-void-circle-conv-M-10}
  \end{subfigure}
  \begin{subfigure}{3in}
    \includegraphics[scale=0.969]{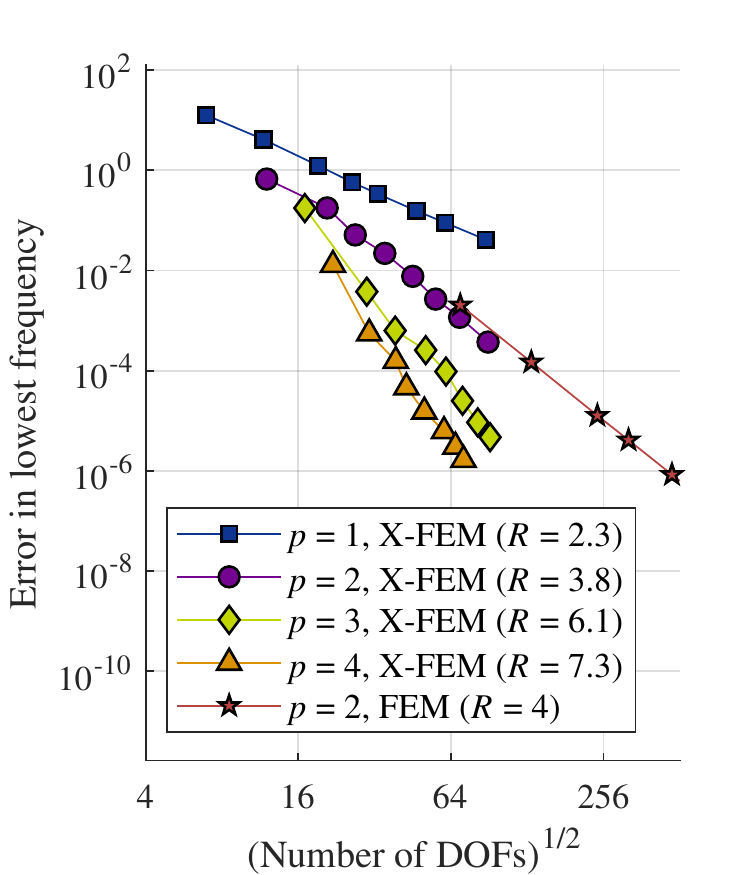}
    \caption{}\label{fig:ex-void-circle-conv-X-1}
  \end{subfigure}
  \begin{subfigure}{3in}
    \includegraphics[scale=0.969]{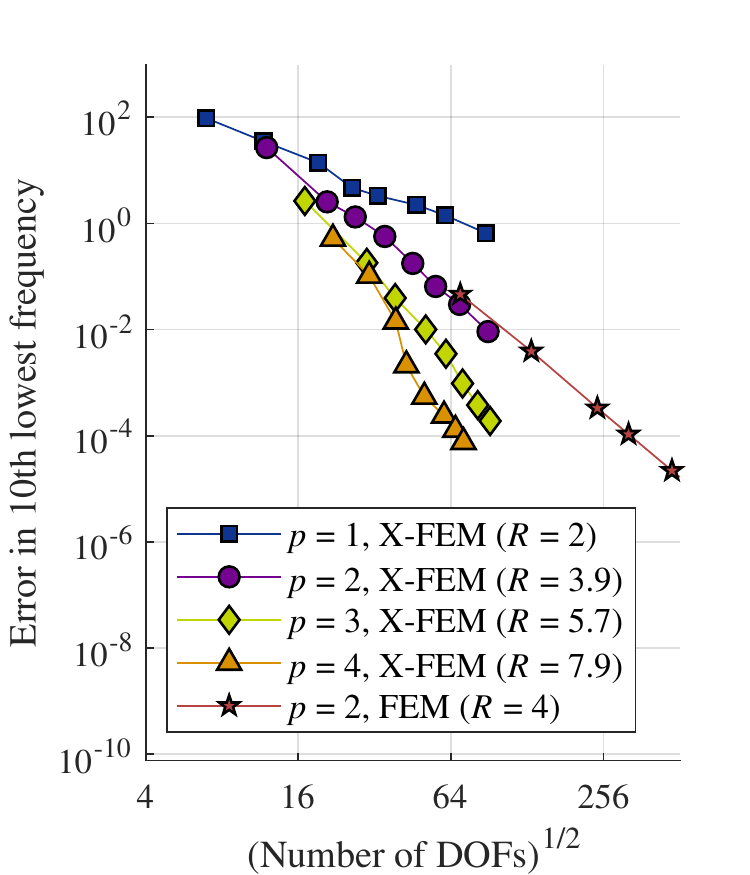}
    \caption{}\label{fig:ex-void-circle-conv-X-10}
  \end{subfigure}
  \caption{Error in frequency versus number of DOFs for the circular void
    example.  (a) Lowest frequency at the X-point, (b) tenth lowest frequency at
    the X-point, (c) lowest frequency at the M-point, and (d) tenth lowest
    frequency at the M-point.}
  \label{fig:ex-void-circle-conv}
\end{figure}

To quantitatively study the accuracy of the extended finite element approach,
the error in frequency as a function of the number of DOFs is plotted in
\fref{fig:ex-void-circle-conv}.  Plots are provided for the error in the lowest
frequency and the tenth-lowest frequency at both the X-point and the M-point. At
the X-point, the reference solution for the lowest frequency is $21.7343273\
\si{\hertz}$ and for the tenth-lowest frequency it is $264.87440\ \si{\hertz}$.
The reference solutions for the M-point are $14.2430873\ \si{\hertz}$ for the
lowest frequency and $226.458845\ \si{\hertz}$ for the tenth-lowest frequency.
Reference solutions are generated from a highly refined FEM mesh of quadratic
triangles, containing 6,285,432 DOFs.  The frequency from the extended finite
element solution approaches the reference solution at a rate of $2p$, matching a
priori error estimates for the FEM.  For quadratic finite elements, a
convergence rate of $4$ is observed in \fref{fig:ex-void-circle-conv}, which
also matches the theoretical estimate. With the spectral X-FEM, accuracy per
degree-of-freedom consistently increases with larger $p$, demonstrating the
solution efficiency gains possible through spectral finite elements.  To
illustrate this point, with quartic extended finite elements, only about 4,000
DOFs are required to obtain accuracy on par with approximately 200,000 DOFs
using a quadratic finite element mesh.

\begin{figure}
  \centering
  \begin{subfigure}{3in}
    \includegraphics[scale=0.969]{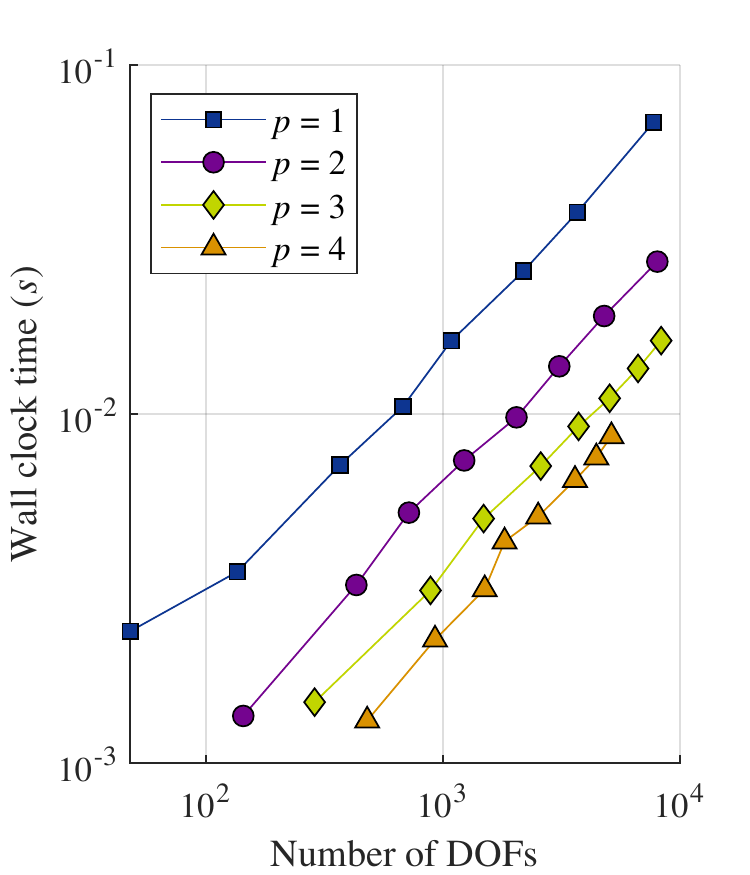}
    \caption{}\label{fig:ex-void-circle-timing-cg}
  \end{subfigure}
  \begin{subfigure}{3in}
    \includegraphics[scale=0.969]{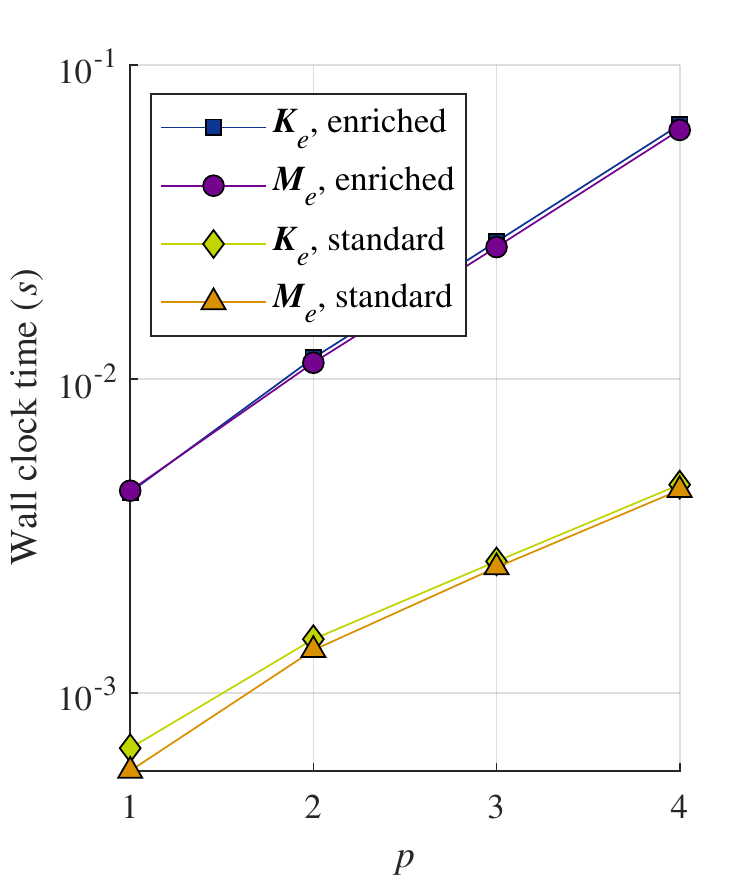}
    \caption{}\label{fig:ex-void-circle-timing-KM}
  \end{subfigure}
  \caption{\revone{Wall clock time for X-FEM related routines in the circular void
    problem. (a) Wall clock time required to determine elements cut by the
    circular void and develop a parametric representation of the circle on each
    cut element versus number of DOFs and (b) wall clock time to integrate
    element system matrices versus $p$ for enriched and standard elements.}}
  \label{fig:ex-void-circle-timing}
\end{figure}

\revone{Solutions generated from the X-FEM require identification of elements
cut by the void boundary, generating the parametric representation over the
element of the boundary, and integrating element stiffness and mass matrices
over the cut element using the HNI method.  To quantify the impact of these
tasks on the analysis run time, we measure the wall clock time required to
perform each of them on each analysis run.  To ensure consistent timing
measurements, each task is repeated 100 times, and the average wall clock time
is reported here.  All timing results are generated on a Linux computer with an
Intel Xeon E5-2695 v4 and 128 GB RAM running MATLAB 9.9.0.  Time to identify cut
elements and generate parametric representations of the boundary in each cut
element versus the number of DOFs is reported in
\fref{fig:ex-void-circle-timing-cg}.  From \fref{fig:ex-void-circle-timing-cg},
we observe that the number of DOFs is linearly correlated to the wall clock time
to complete these tasks.  Since fewer elements per DOF are present as $p$
increases, fewer parametric representations of the circle need to be generated
as $p$ grows.  As a result, the wall clock time per DOF reduces as $p$
increases.  The time to integrate the element stiffness and mass matrices
($\vm{K}_e$ and $\vm{M}_e$, respectively) in enriched and standard elements
versus $p$ is displayed in \fref{fig:ex-void-circle-timing-KM}. Overall,
integrating element system matrices using HNI on cut elements increases the wall
clock time by approximately a factor of 10.  While the increase in time to
integrate the element system matrices using HNI is considerable, we note
integration remains a small portion of total wall clock time to compute the band
structure.  Furthermore, the additional time required to compute void boundary
parametrizations and to integrate using the HNI method is in lieu of generating
a conforming mesh---a nontrivial task that is required for the finite element
solution.  Finally, the performance of the MATLAB routines could likely be
improved through further code optimizations, which were not investigated.}

\subsection{Phononic crystal with circular void and skew domain}
\begin{figure}
  \centering
  \begin{subfigure}{3.15in}
    \includegraphics[scale=0.969]{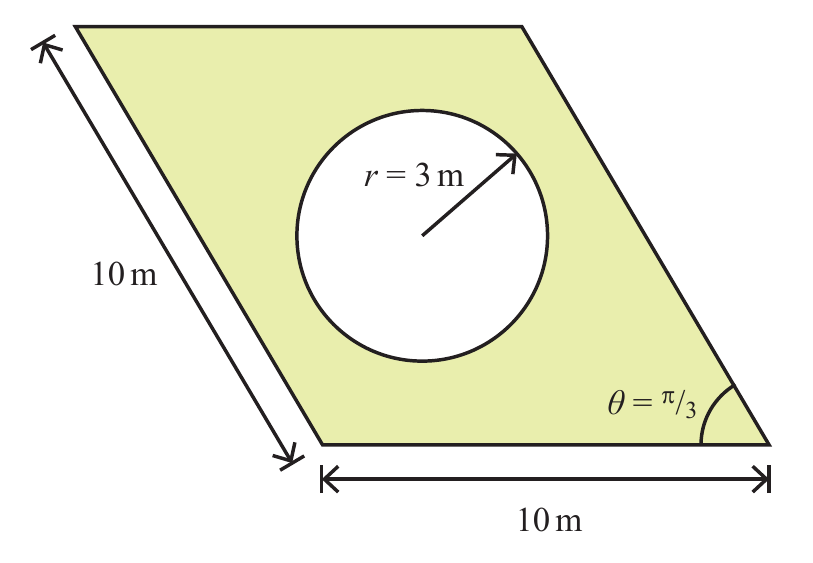}
    \caption{}\label{fig:ex-void-skewcircle-setup}
  \end{subfigure}
  \begin{subfigure}{3.15in}
    \includegraphics[scale=0.969]{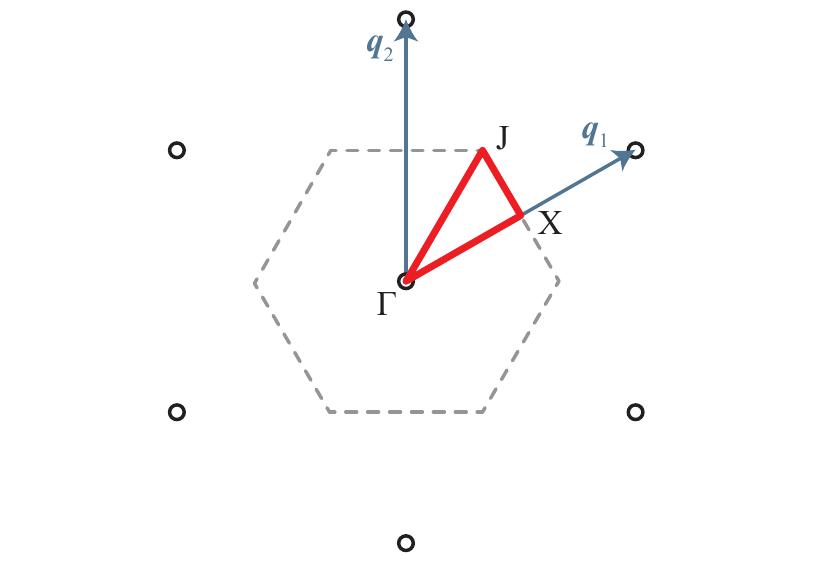}
    \caption{}\label{fig:ex-void-skewcircle-ibz}
  \end{subfigure}
  \begin{subfigure}{3.15in}
    \includegraphics[scale=0.969]{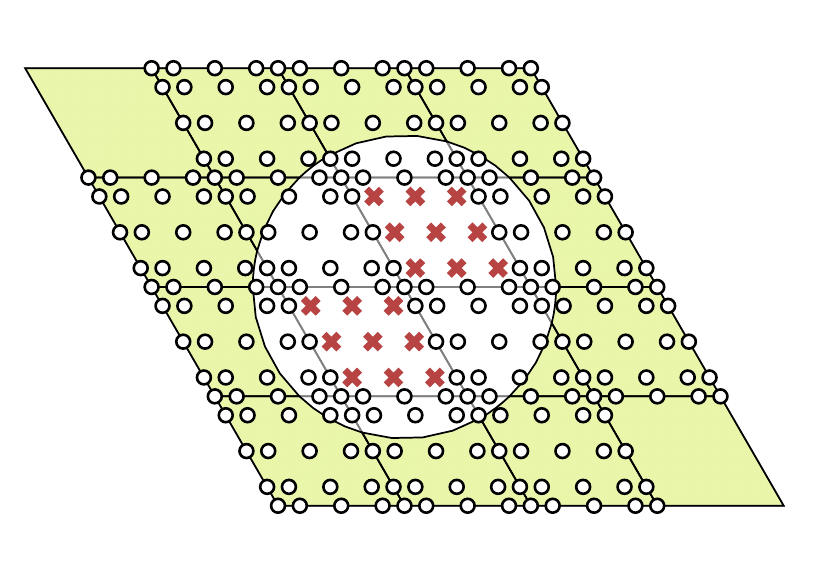}
    \caption{}\label{fig:ex-void-skewcircle-mesh-xfem}
  \end{subfigure}
  \begin{subfigure}{3.15in}
    \includegraphics[width=3.15in]{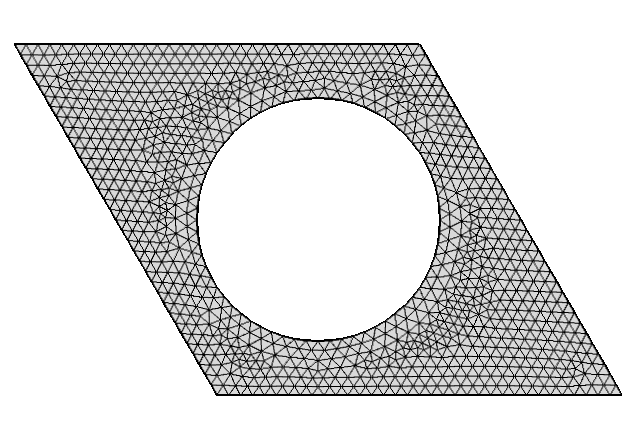}
    \caption{}\label{fig:ex-void-skewcircle-mesh-fem}
  \end{subfigure}
  \begin{subfigure}{4.5in}
    \centering
    \includegraphics[scale=0.969]{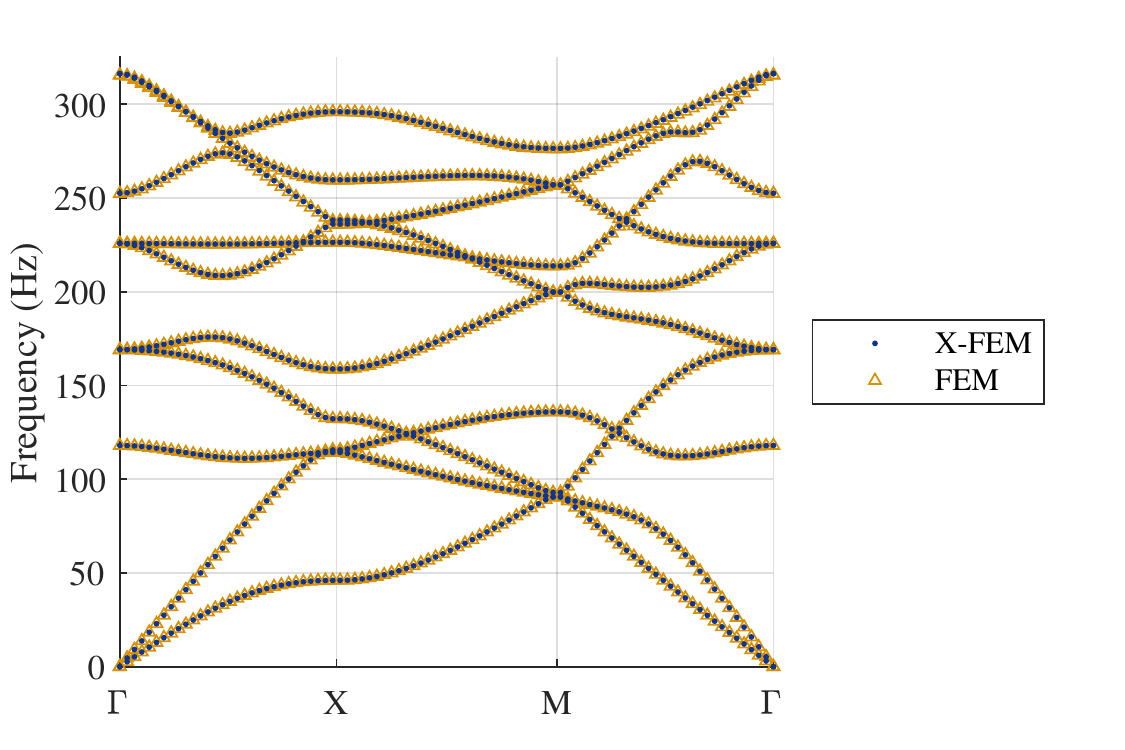}
    \caption{}\label{fig:ex-void-skewcircle-bsd}
  \end{subfigure}
  \caption{Skewed unit cell geometry of a phononic crystal with a circular void.
    (a) Problem setup, (b) first Brillouin zone on the reciprocal lattice
    (dashed line) and irreducible Brillouin zone (bold, red line), (c) extended
    finite element mesh (476 total DOFs; nodes with active DOFs are open circles
    and nodes with removed DOFs are red crosses), (d) finite element mesh (7,632
    DOFs), and (e) band structure diagram.}
  \label{fig:ex-void-skewcircle}
\end{figure}

In this example, the band structure of a skew unit cell with a circular void is
generated.  The geometry of the unit cell is illustrated in
\fref{fig:ex-void-skewcircle-setup} and the material properties are repeated
from \sref{ssec:ex-hole-circle}; that is, $E = 20\ \si{\giga\pascal}$, $\nu =
0.25$, and $\rho = 2700\ \si{\kilo\gram\per\meter\cubed}$.  This choice of
lattice vectors results in a regular hexagonal first Brillouin zone, which is
illustrated in \fref{fig:ex-void-skewcircle-ibz}.

The band structure diagram is computed using both the FEM and the X-FEM.  For
the X-FEM, the problem is run with a $4 \times 4$ element mesh of structured
quartic elements, resulting in a total of 476 DOFs.  The mesh is shown in
\fref{fig:ex-void-skewcircle-mesh-xfem}.  As is done in the example in
\sref{ssec:ex-hole-circle}, the circular void is modeled exactly using a
quadratic rational B\'{e}zier curve. For the FEM, a mesh of quadratic triangular
elements is utilized with a total of 7,632 DOFs (see
\fref{fig:ex-void-skewcircle-mesh-fem}).  With the FEM, the circular void is
only approximated using quadratic polynomial curves.  Band structure for the
finite element and extended finite element solutions is presented in
\fref{fig:ex-void-skewcircle-bsd}.  Quantitatively, relative errors between the
band structures range from $4.7 \times 10^{-5}$ to $2.7 \times 10^{-3}$.

\subsection{Phononic crystal with elliptical void}
\begin{figure}[t!]
  \centering
  \begin{subfigure}{3in}
    \includegraphics[scale=0.969]{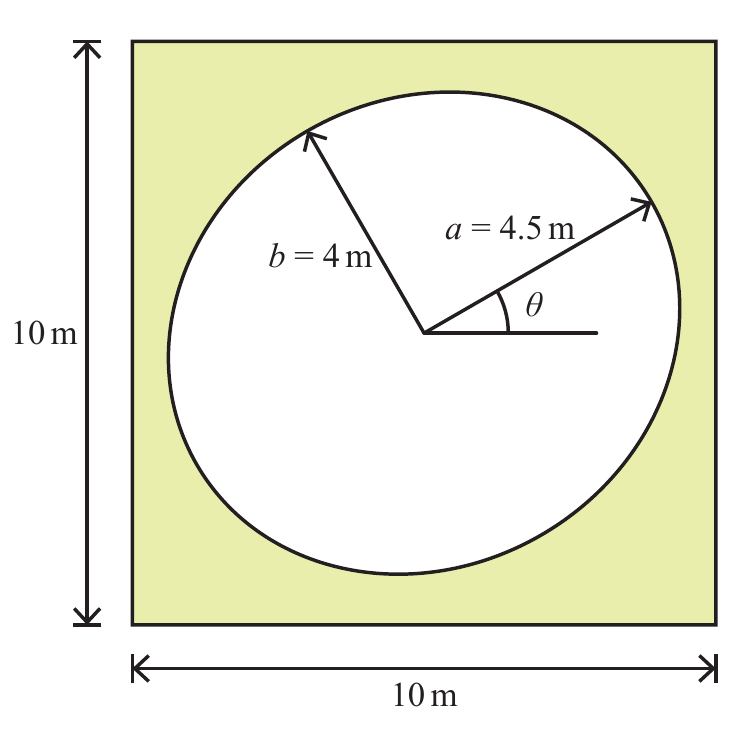}
    \caption{}\label{fig:ex-void-ellipse-setup}
  \end{subfigure}
  \begin{subfigure}{3in}
    \includegraphics[scale=0.969]{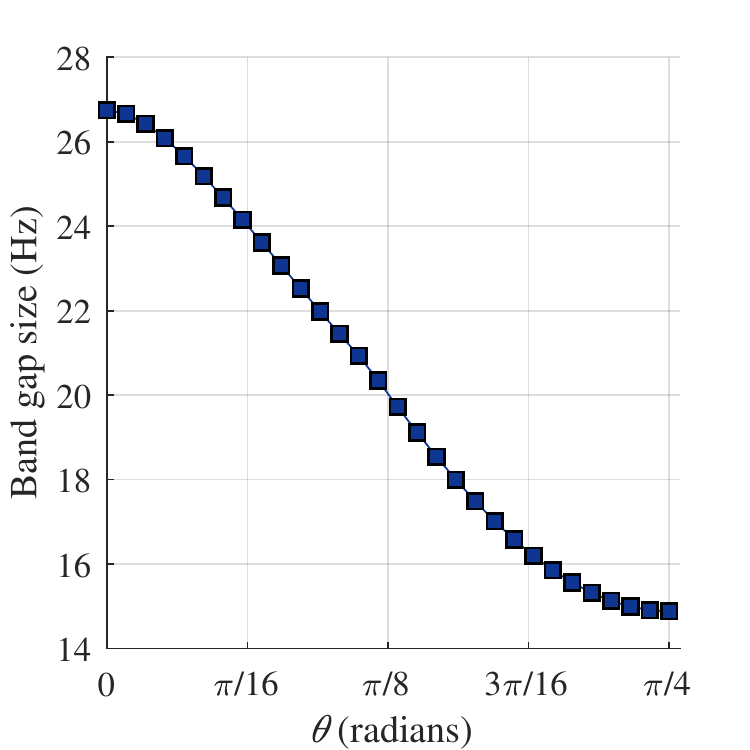}
    \caption{}\label{fig:ex-void-ellipse-bgvstheta}
  \end{subfigure}
  \begin{subfigure}{3in}
    \includegraphics[scale=0.969]{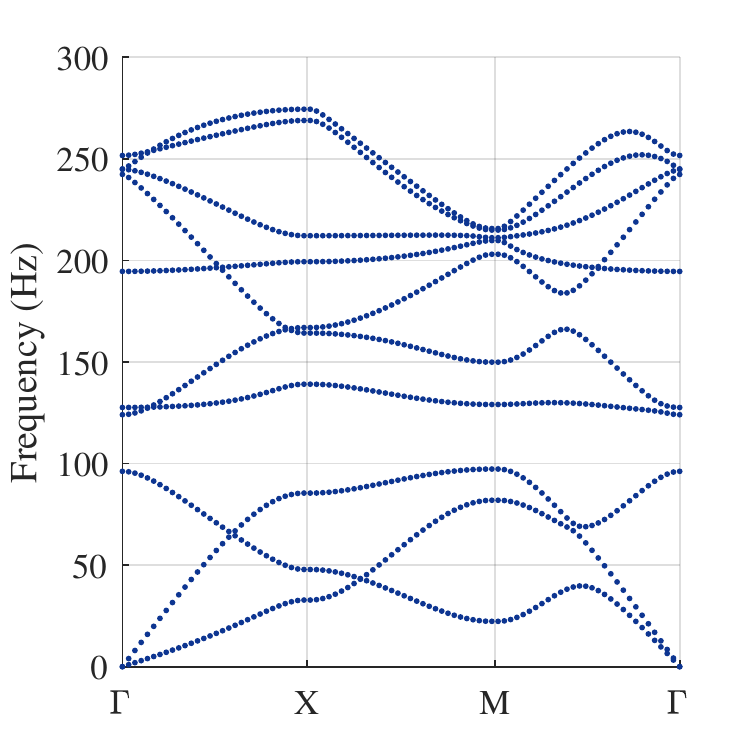}
    \caption{}\label{fig:ex-void-ellipse-th0-bsd-xfem}
  \end{subfigure}
  \begin{subfigure}{3in}
    \includegraphics[scale=0.969]{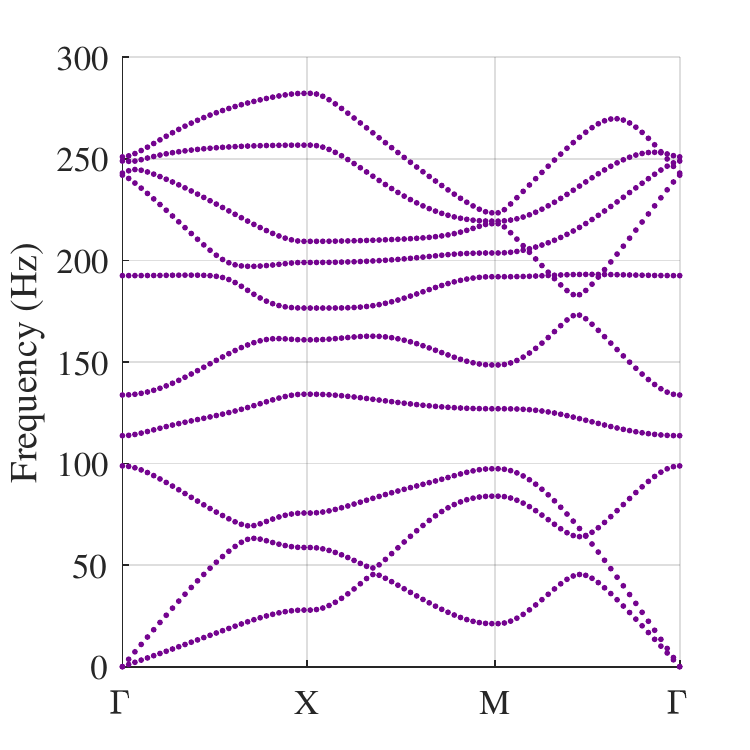}
    \caption{}\label{fig:ex-void-ellipse-th45-bsd-xfem}
  \end{subfigure}
  \caption{Unit cell geometry of a phononic crystal with an elliptical void. (a)
    Problem setup, (b) largest band gap versus angle $\theta$, (c) extended
    finite element band structure diagram for $\theta = 0$, and (d) extended
    finite element band structure diagram for $\theta = \pi / 4$.}
  \label{fig:ex-void-ellipse}
\end{figure}

Next, a band structure diagram is constructed for a phononic crystal whose
Bloch-periodic domain contains an elliptical void.  The material properties from
\sref{ssec:ex-hole-circle} are repeated in this example and the unit cell
geometry is illustrated in \fref{fig:ex-void-ellipse-setup}.  The parameter
$\theta$ is varied from $0$ to $\pi / 4$ and the resulting size of the largest
band gap in the first ten frequencies is plotted in
\fref{fig:ex-void-ellipse-bgvstheta}.  Two representative band structure
diagrams are presented in \fref{fig:ex-void-ellipse-th0-bsd-xfem} and
\fref{fig:ex-void-ellipse-th45-bsd-xfem}, which are for $\theta = 0$ and $\theta
= \pi / 4$, respectively.  When $\theta = 0$, a single band gap is present
between approximately $100\ \si{\hertz}$ and $125\ \si{\hertz}$.  When $\theta$
is increased to $\pi / 4$, two band gaps are present at approximately $100\
\si{\hertz}$ and $175\ \si{\hertz}$.  All extended finite element analyses of
this problem are performed with a $4 \times 4$ element mesh of quartic finite
elements.  The ellipse geometry is modeled exactly using rational quadratic
B\'{e}zier curves to represent the void boundary.

\subsection{Phononic crystal with a clover-shaped void}
\begin{figure}[t!]
  \centering
  \begin{subfigure}{3in}
    \includegraphics[scale=0.969]{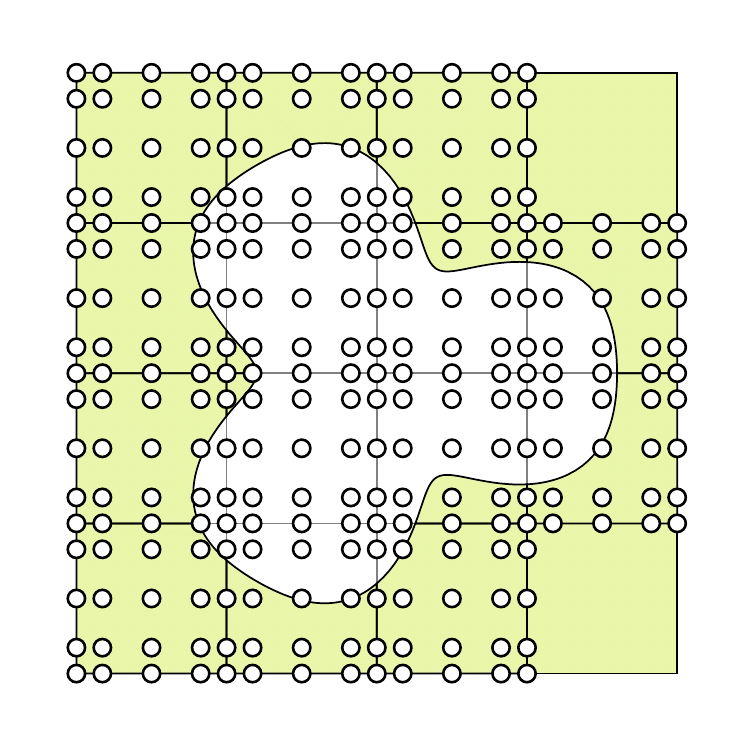}
    \caption{}\label{fig:ex-void-wavycircle-mesh}
  \end{subfigure}
  \begin{subfigure}{3in}
    \centering
    \includegraphics[scale=0.51]{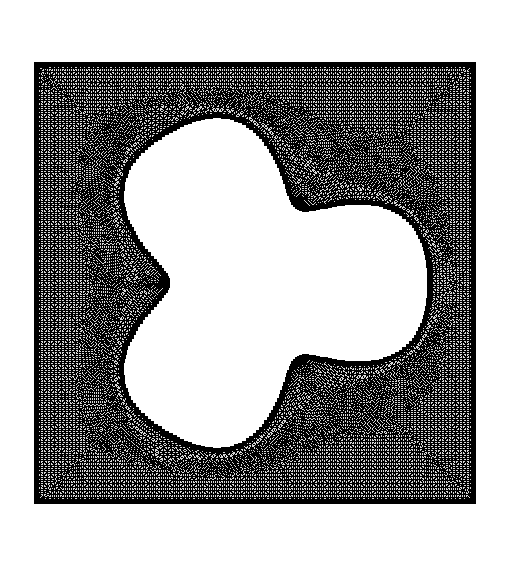}
    \caption{}\label{fig:ex-void-wavycircle-mesh-fem}
  \end{subfigure}
  \begin{subfigure}{4.5in}
    \centering
    \includegraphics[scale=0.969]{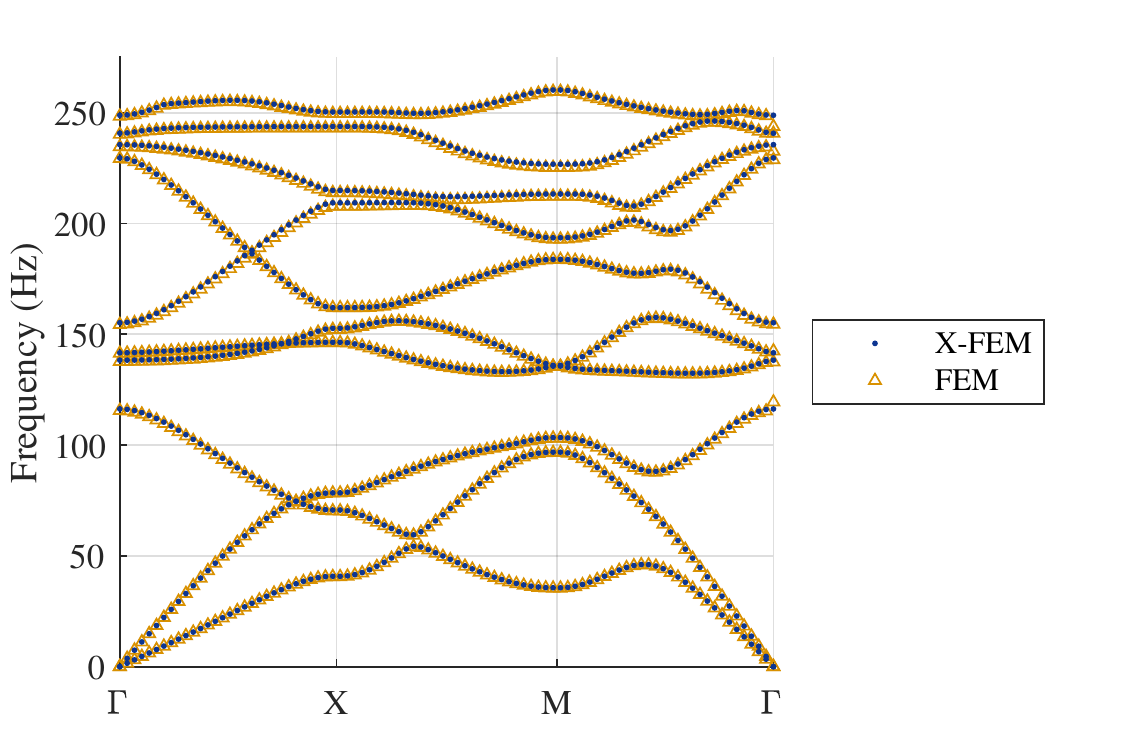}
    \caption{}\label{fig:ex-void-wavycircle-bsd}
  \end{subfigure}
  \caption{Unit cell geometry of a phononic crystal with a clover-shaped void.
    (a) extended finite element mesh (512 total DOFs; nodes with enriched DOFs
    are open circles), \revone{(b) finite element mesh (74,204 DOFs),} and (c)
    band structure diagram.}
  \label{fig:ex-void-wavycircle}
\end{figure}

As the final void example, we consider a unit cell $\Omega = [-5,5]^2$ with a
void defined by a level set function, which demonstrates the capability to model more
general voids whose geometry may not be known explicitly.  We choose the level
set function
\begin{equation}\label{eq:ls-wavycircle}
  \varphi(r, \theta) = \frac{1}{5} r + \frac{2}{5} \sin^4 \left( 
    \frac{3}{2} \theta
   \right) - \frac{4}{5} ,
\end{equation}
where $r = \sqrt{x^2 + y^2}$ and $\theta = \tan^{-1} \left(\frac{y}{x}\right)$
are polar coordinates.  The level set boundary is approximated using cubic
Hermite functions, as described in \sref{ssec:level-set}.  Rather than
approximating the level set function using the finite element interpolant, we
choose to use \eqref{eq:ls-wavycircle} directly in \eqref{eq:objective-fn} and
\eqref{eq:ls-approx-error}.  This results in the cubic Hermite functions
approximating the level set function directly, eliminating a source of error in
the geometric description of the unit cell.  The domain is modeled with 16
quartic extended finite elements arranged in a $4 \times 4$ mesh.  Material
properties match those in \sref{ssec:ex-hole-circle}.  The meshed geometry is
illustrated in \fref{fig:ex-void-wavycircle-mesh} and the resulting band
structure for the ten lowest frequencies is displayed in
\fref{fig:ex-void-wavycircle-bsd}.  \revone{The resulting band structure is in
close agreement with one generated from a finite element analysis using
quadratic triangular elements (mesh drawn in
\fref{fig:ex-void-wavycircle-mesh-fem}).   A maximum difference of $0.6\
\si{\hertz}$ is observed between the finite element and extended finite element
band structure diagrams.}

\subsection{One-dimensional two-phase phononic crystal}
First, the application of the enrichment function for modeling material
discontinuities in the X-FEM is presented in a one-dimensional setting.  In one
dimension, Rytov~\cite{rytov1956acoustical} derived an exact dispersive
solution:
\begin{subequations}\label{eq:rytov-soln}
\begin{equation}
  \cos (ka) = \cos \frac{\omega h_1}{c_1} \cos \frac{\omega h_2}{c_2}
    - \Gamma \sin \frac{\omega h_1}{c_1} \sin \frac{\omega h_2}{c_2} ,
\end{equation}
where
\begin{equation}
  \Gamma = \frac{1+\kappa^2}{2\kappa}, \qquad \kappa = \frac{\rho_1 c_1}{\rho_2 c_2},
  \qquad c_i = \sqrt{\frac{E_i}{\rho_i}} \quad (i = 1, 2).
\end{equation}
\end{subequations}
Following Lu and Srivastava~\cite{lu2016variational}, we choose the material
properties $E_1 = 8\ \si{\giga\pascal}$, $\rho_1 = 1000\
\si{\kilo\gram\per\meter\cubed}$, $h_1 = 0.003\ \si{\meter}$, $E_2 = 300\
\si{\giga\pascal}$, $\rho_2 = 8000\ \si{\kilo\gram\per\meter\cubed}$, and $h_2 =
0.0013\ \si{\meter}$.  The unit cell length is $a = h_1 + h_2 =
0.0043\ \si{\meter}$.  Given a frequency $f = \omega / (2 \pi)$,
\eqref{eq:rytov-soln} can be used to find a wavevector in the irreducible
Brillouin zone, $Q := k a / (2 \pi)$ for $-0.5 \leq Q \leq 0.5$ (or $0 \leq Q
\leq 0.5$ due to symmetry).

The problem is modeled using the FEM, the X-FEM, and the plane wave expansion
method.  For analyses conducted with finite elements, a three element mesh is
used where two elements lie in the domain of material 1 and the domain of
material 2 is modeled with one element. For extended finite element analyses,
the mesh consists of two elements, and the inter-element node does not coincide
with the location of the interface.  Instead, \eqref{eq:bimat-enrichment} is
used to capture the effect of the material interface in the interior of the
element. The accuracy of the X-FEM and the FEM for these meshes is expected to
be nearly identical provided $p$ is the same, since both are capable of
representing the same piecewise continuous polynomial
functions~\cite{Chin:2019:MCI}.

\begin{figure}
  \centering
  \begin{subfigure}{3.15in}
    \includegraphics[scale=0.969]{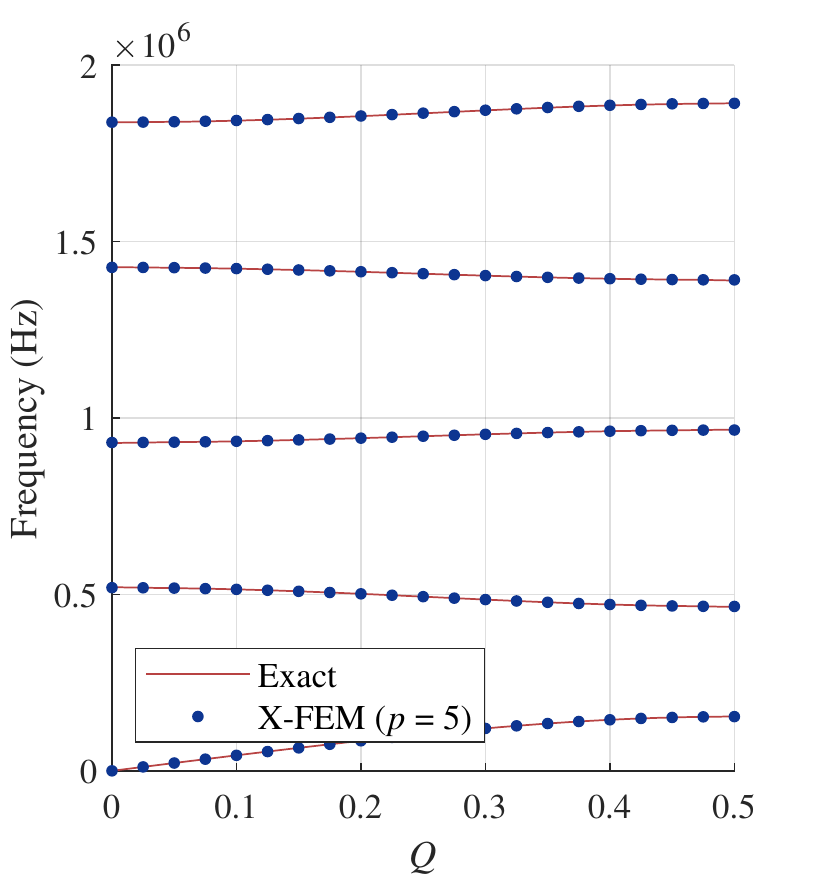}
    \caption{}\label{fig:ex-bimat-rytov-bsd-xfem}
  \end{subfigure}
  \begin{subfigure}{3.15in}
    \includegraphics[scale=0.969]{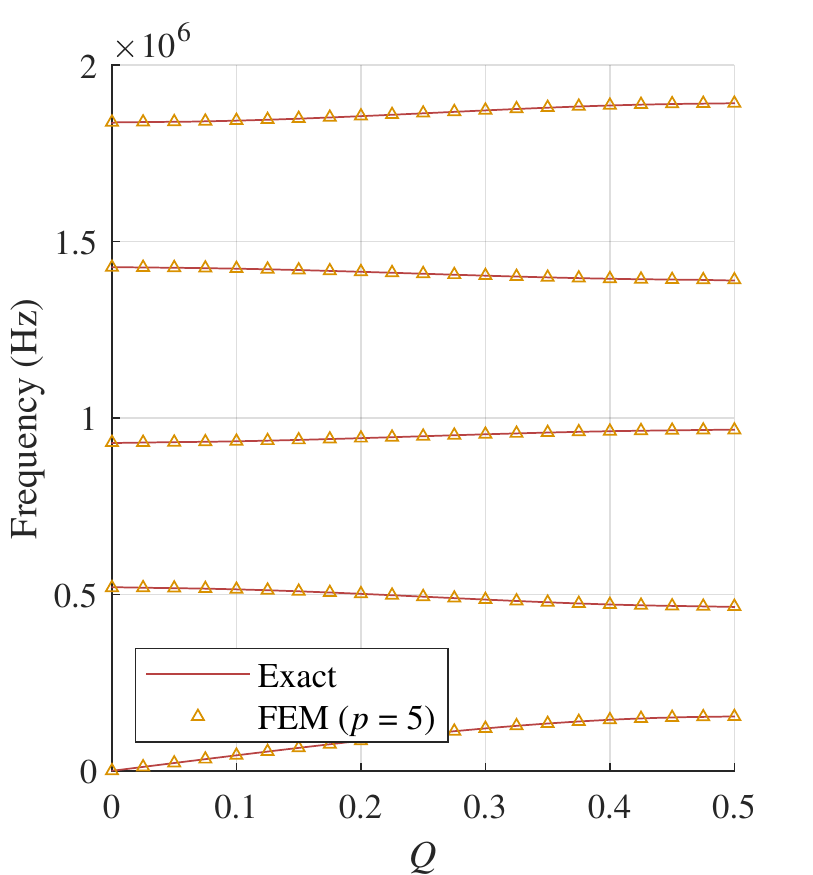}
    \caption{}\label{fig:ex-bimat-rytov-bsd-fem}
  \end{subfigure}
  \begin{subfigure}{3.15in}
    \includegraphics[scale=0.969]{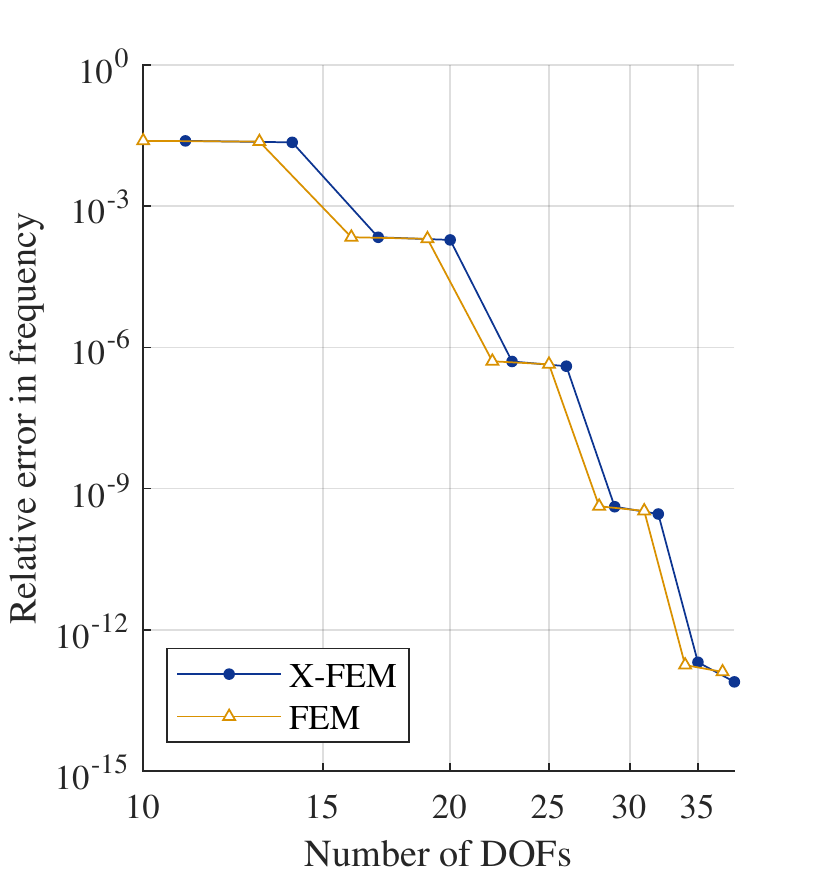}
    \caption{}\label{fig:ex-bimat-rytov-pconv-q0}
  \end{subfigure}
  \begin{subfigure}{3.15in}
    \includegraphics[scale=0.969]{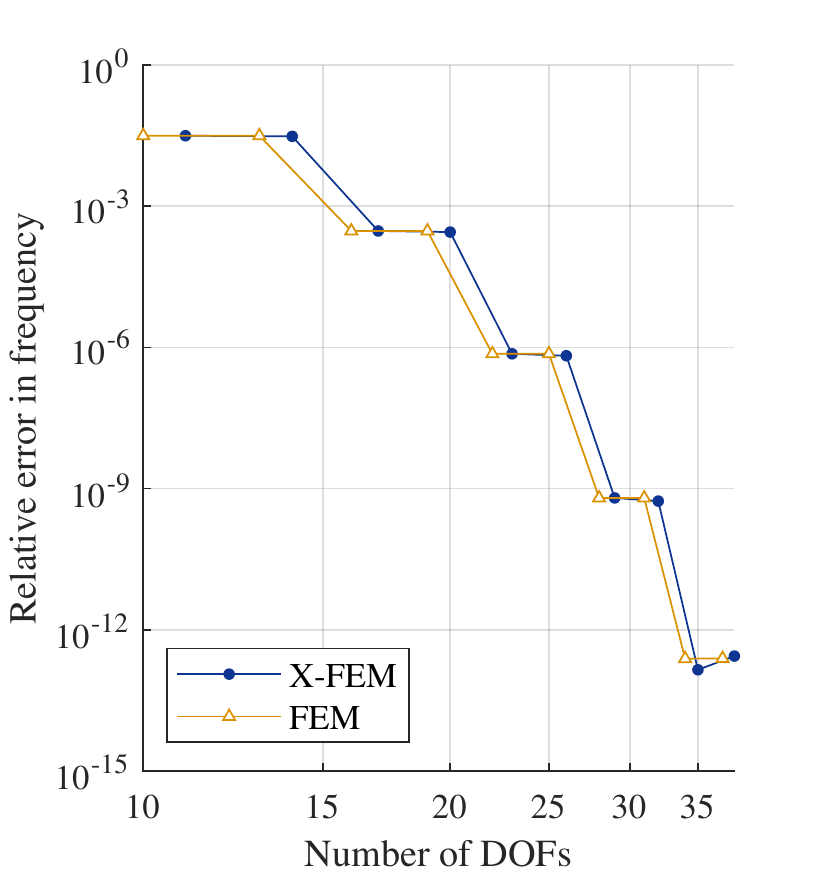}
    \caption{}\label{fig:ex-bimat-rytov-pconv-q0p5}
  \end{subfigure}
  \caption{One-dimensional two-phase phononic crystal.  (a) Band structure
    diagram using two fifth-order extended finite elements, (b) band structure
    diagram using three fifth-order finite elements, (c) convergence of the
    fifth-lowest frequency with $p$-refinement at $Q=0$, and (d) convergence of
    the fifth-lowest frequency with $p$-refinement at $Q=0.5$.}
  \label{fig:ex-bimat-rytov}
\end{figure}

To examine the effect of $p$-refinement on solution accuracy, the problem is run
with finite elements and extended finite elements of order $3$ to $12$.  The
band structure diagram for the case $p=5$ is presented in
\fref{fig:ex-bimat-rytov-bsd-fem} and \fref{fig:ex-bimat-rytov-bsd-xfem}.  This
results in 16 DOFs for the finite element analysis and 17 DOFs for the extended
finite element analysis.  For the five lowest frequencies, the numerical
solutions are indistinguishable from the exact band structure diagram using
either the FEM or the X-FEM.  Relative frequency error in the fifth-lowest
frequency versus the number of degrees of freedom is plotted in
\fref{fig:ex-bimat-rytov-pconv-q0} and \fref{fig:ex-bimat-rytov-pconv-q0p5} for
$Q=0$ and $Q=0.5$, respectively.  Convergence of both the finite element
solution and the extended finite element solution is exponential with
$p$-refinement.

\begin{figure}
  \centering
  \begin{subfigure}{3in}
    \includegraphics[scale=0.969]{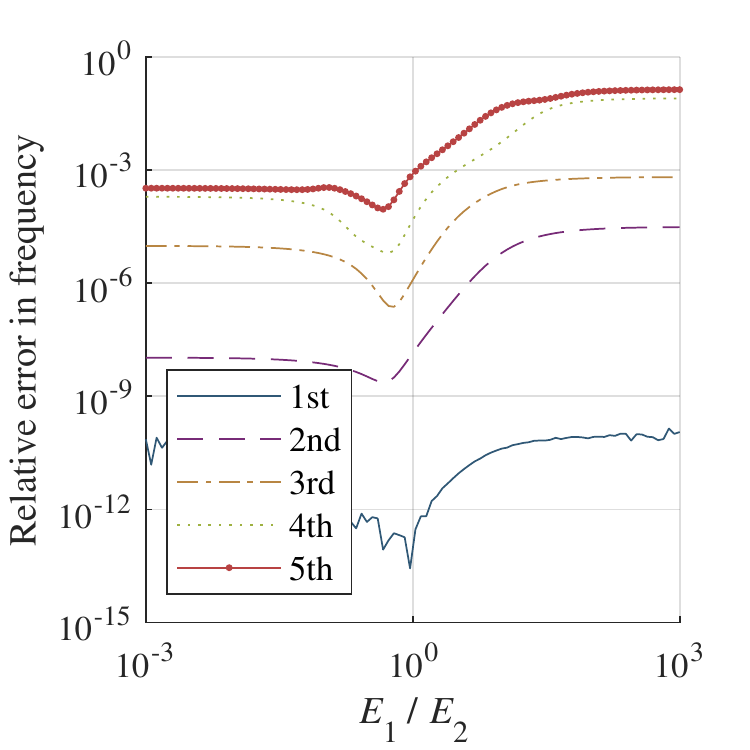}
    \caption{}\label{fig:ex-bimat-rytov-ediff-xfem}
  \end{subfigure}
  \begin{subfigure}{3in}
    \includegraphics[scale=0.969]{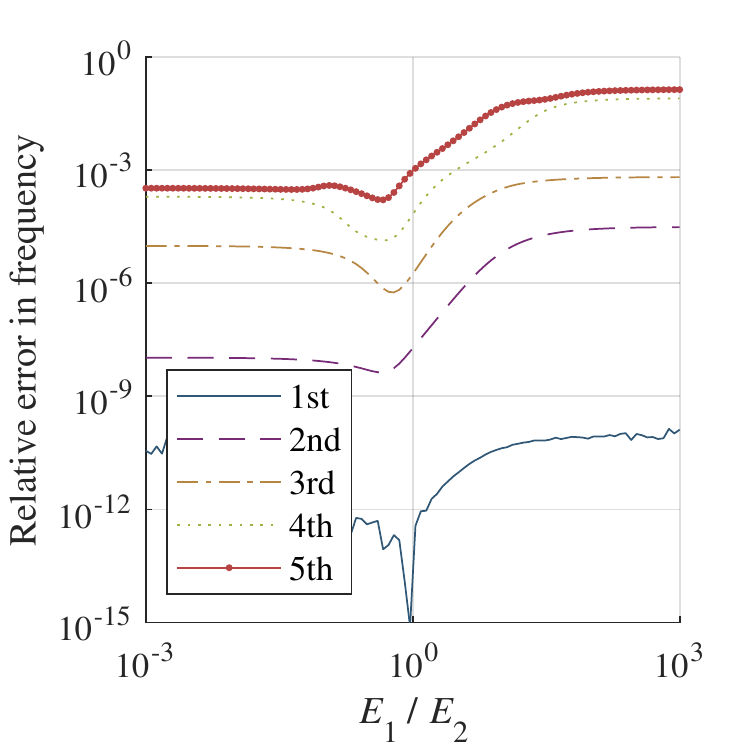}
    \caption{}\label{fig:ex-bimat-rytov-ediff-fem}
  \end{subfigure}
  \begin{subfigure}{3in}
    \includegraphics[scale=0.969]{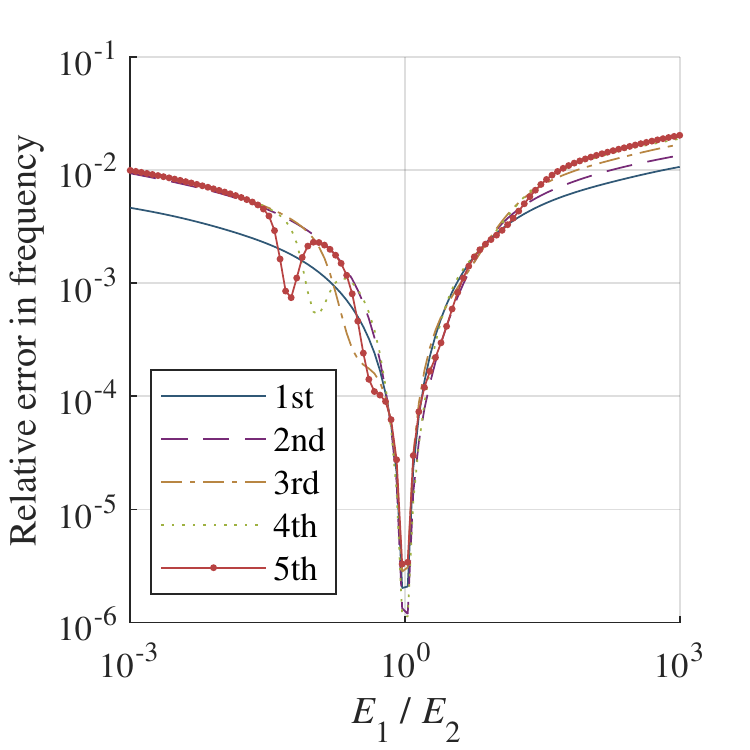}
    \caption{}\label{fig:ex-bimat-rytov-ediff-pwe}
  \end{subfigure}
  \caption{Error in relative frequency for the first through fifth lowest
    frequency for various ratios of $E_1$ to $E_2$.  (a) Fifth-order extended
    finite elements (17 DOFs), (b) fifth-order spectral finite elements (16
    DOFs), (c) plane wave expansion (201 DOFs).}
  \label{fig:ex-bimat-rytov-ediff}
\end{figure}

Finally, we examine the accuracy of the frequency solutions as a function of the
ratio of $E_1$ to $E_2$.  A total of 100 values of $E_1 / E_2$ in between $E_1 /
E_2 = 1.0 \times 10^{-3}$ to $E_1 / E_2 = 1.0 \times 10^3$ are tested in the
numerical study, and the error in the lowest five frequencies as $E_1 / E_2$ is
varied is plotted at the point $Q = 0.25$ in \fref{fig:ex-bimat-rytov-ediff}.
The study is conducted with fifth-order spectral finite elements (16 DOFs),
fifth-order extended spectral finite elements (17 DOFs), and using the plane
wave expansion method (201 DOFs).  Figure~\ref{fig:ex-bimat-rytov-ediff}
demonstrates superior accuracy of spectral (extended) finite elements as
compared to the plane wave expansion technique in two-phase materials.  With a
greater than $10$-fold reduction of DOFs, the FEM and the X-FEM deliver much
better accuracy than the plane wave expansion method for the three lowest
frequencies, and similar accuracy for the fourth and fifth frequency.  The
extended finite element solution contains slightly less error when compared to
the finite element solution due to the extra degree of freedom in the extended
finite element mesh, which allows $(p\!+\!1)$-th order terms to be partially
captured (see Chin and Sukumar~\cite{Chin:2019:MCI} for details).

\subsection{Two-phase phononic crystal with circular inclusion}

\begin{figure}
  \centering
  \includegraphics[scale=0.969]{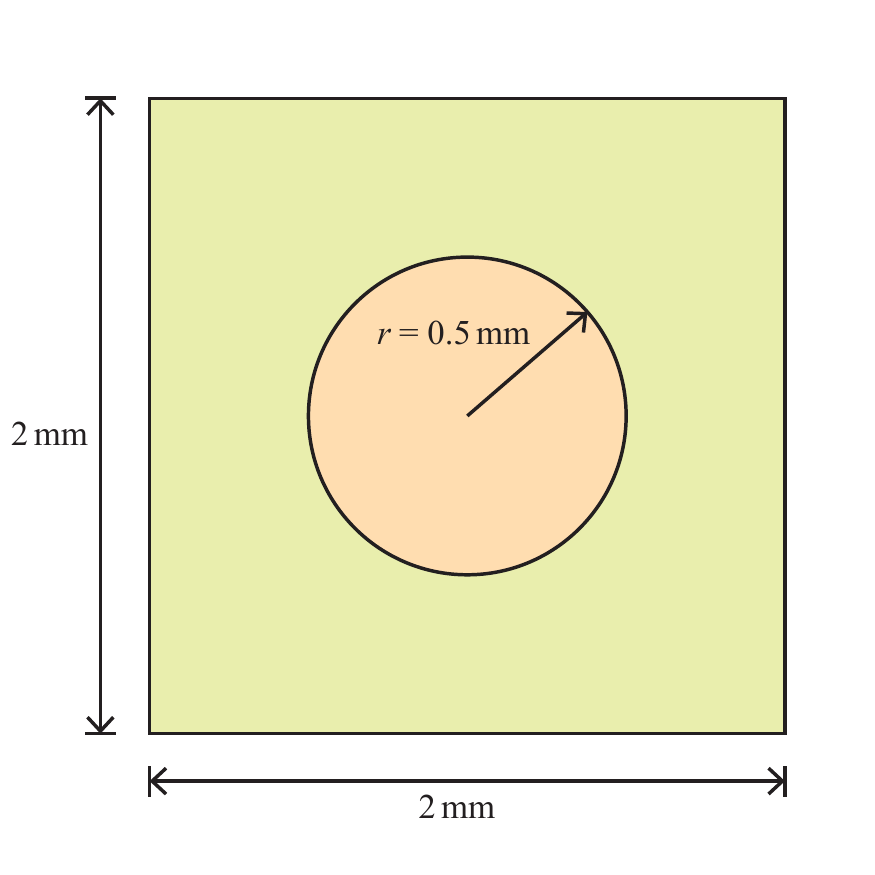}
  \caption{Unit cell geometry of a phononic crystal with a circular inclusion.}
  \label{fig:ex-bimat-circle-setup}
\end{figure}

To study the performance of the enrichment function in a two-dimensional
Bloch-periodic domain, we model a two-phase phononic crystal with a circular
inclusion.  The model problem, which appears in Srivastava and
Nemat-Nasser~\cite{srivastava2014mixed}, is illustrated in
\fref{fig:ex-bimat-circle-setup}.  Inside the circular inclusion, the material
properties are $E_2 = 200\ \si{\giga\pascal}$, $\nu_2 = 0.3$ and $\rho_2 = 3300\
\si{\kilo\gram\per\meter\cubed}$.  Over the remainder of the domain, the
material properties are $E_1 = 2\ \si{\giga\pascal}$, $\nu_1 = 0.3$, and $\rho_1
= 1100\ \si{\kilo\gram\per\meter\cubed}$.  The problem is analyzed using
quadratic, triangular finite elements and the spectral X-FEM.  Since enrichment
using \eqref{eq:bimat-enrichment} requires a finite element interpolation of
level set geometry, the parametric form of the interface location is
approximated using cubic Hermite functions (see \sref{ssec:level-set} for
details).  Reference solutions are generated using a finite element solution
from a highly refined mesh (1,772,522 DOFs).

\begin{figure}
  \centering
  \includegraphics[scale=0.969]{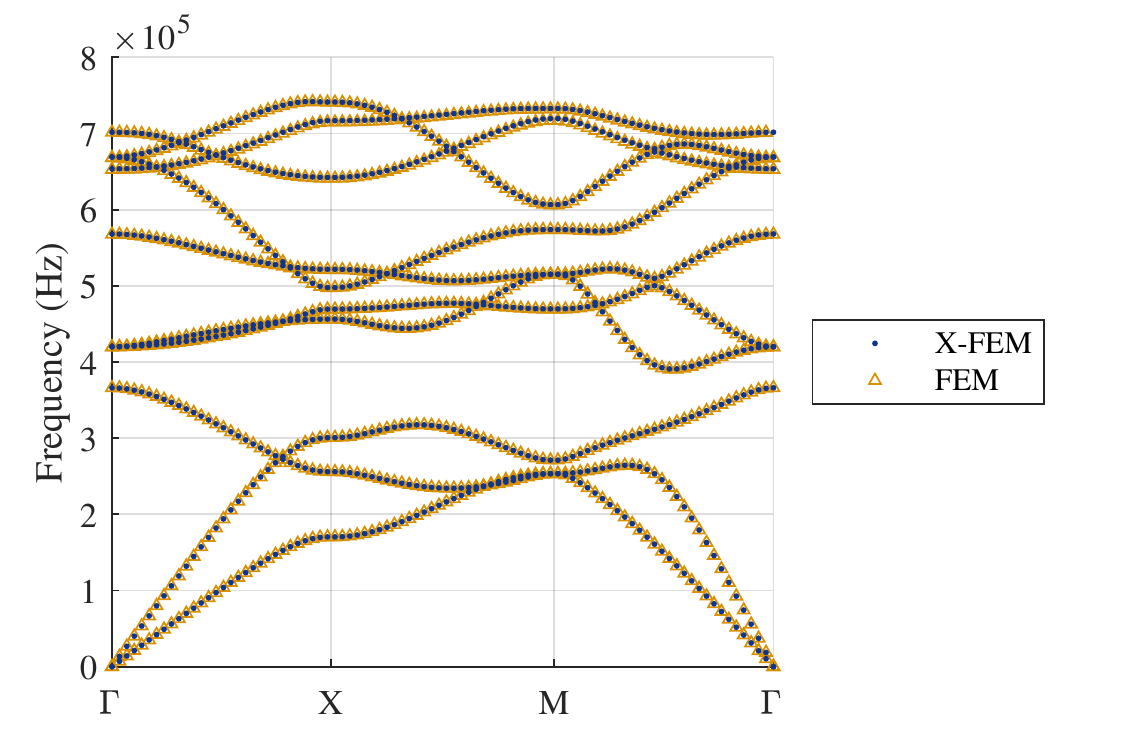}
  \caption{Band structure diagram for a phononic crystal with a circular
    inclusion. Solution using quartic extended finite elements (674 DOFs) and
    reference solution using quadratic finite elements (450,370 DOFs).}
  \label{fig:ex-bimat-circle-bsd}
\end{figure}

\begin{figure}
  \centering
  \begin{subfigure}{3in}
    \includegraphics[scale=0.969]{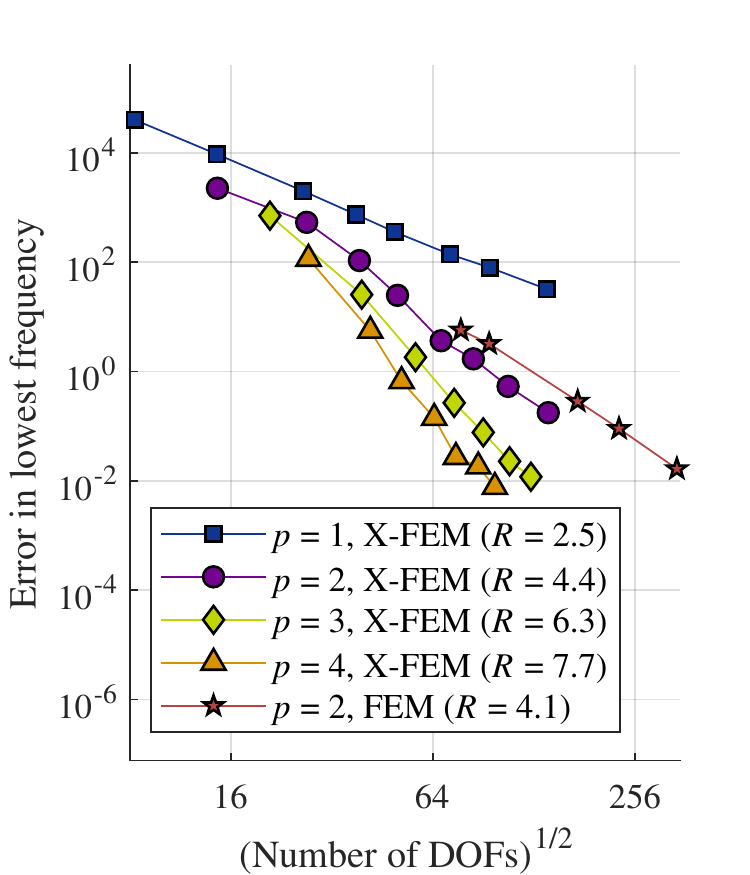}
    \caption{}\label{fig:ex-bimat-circle-conv-M-1}
  \end{subfigure}
  \begin{subfigure}{3in}
    \includegraphics[scale=0.969]{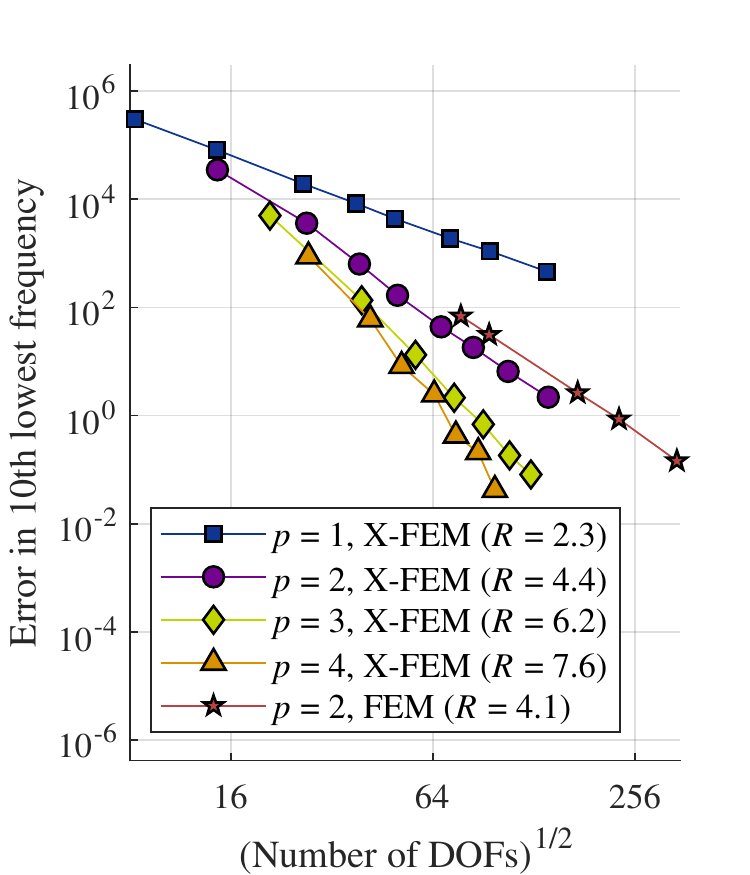}
    \caption{}\label{fig:ex-bimat-circle-conv-M-10}
  \end{subfigure}
  \begin{subfigure}{3in}
    \includegraphics[scale=0.969]{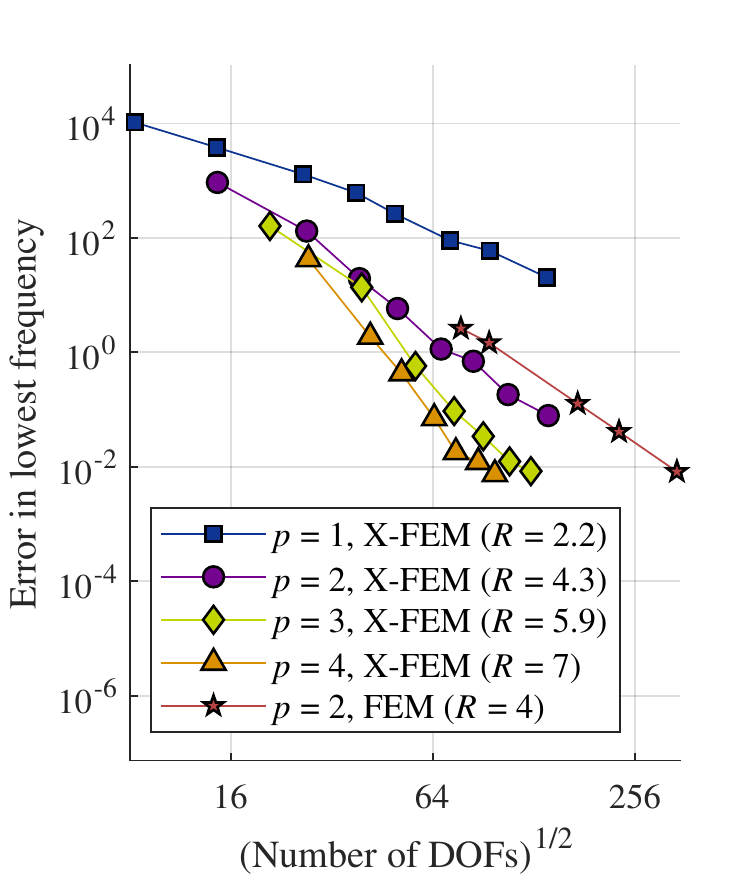}
    \caption{}\label{fig:ex-bimat-circle-conv-X-1}
  \end{subfigure}
  \begin{subfigure}{3in}
    \includegraphics[scale=0.969]{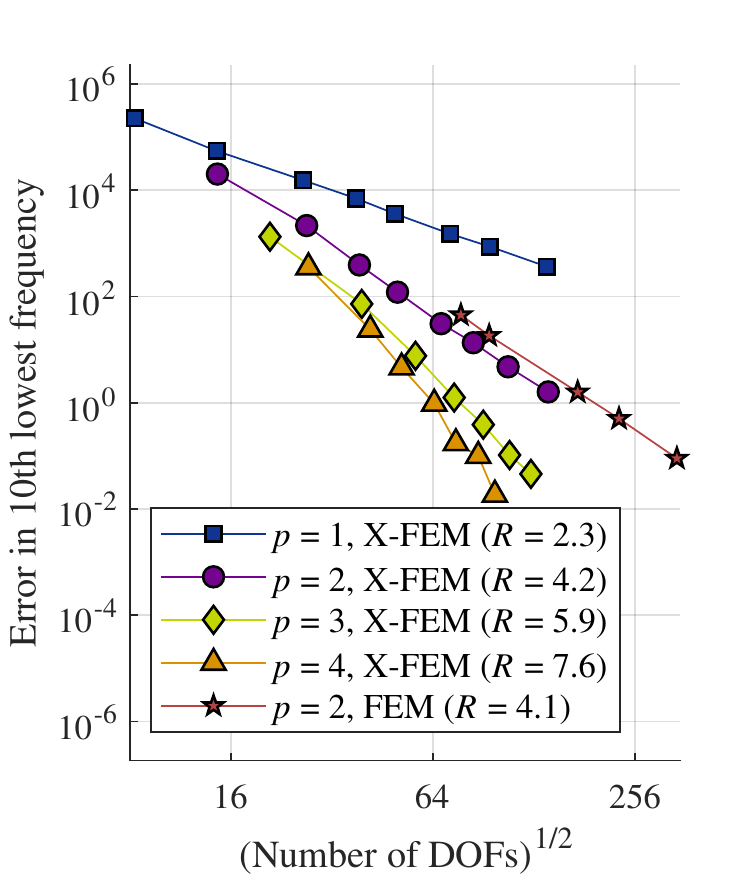}
    \caption{}\label{fig:ex-bimat-circle-conv-X-10}
  \end{subfigure}
  \caption{Error in frequency versus number of DOFs for the circular inclusion
  example.  (a) Lowest frequency at the X-point, (b) tenth lowest frequency at
  the X-point, (c) lowest frequency at the M-point, and (d) tenth lowest
  frequency at the M-point.}
  \label{fig:ex-bimat-circle-conv}
\end{figure}

A band structure diagram for this problem using both finite elements and
spectral extended finite elements is presented in
\fref{fig:ex-bimat-circle-bsd}.  The extended finite element solution using a $4
\times 4$ mesh of quartic finite elements (674 DOFs) is indistinguishable from
the reference finite element solution.  The convergence properties of both the
spectral extended finite element and finite element solutions are examined
through $h$-refinement studies in \fref{fig:ex-bimat-circle-conv}.  Spectral
extended finite elements of order $p=1, \dotsc, 4$ and quadratic finite elements
are utilized in the study.  The error in the lowest and tenth-lowest frequency
at the X-point and the M-point are considered.  Reference solutions (computed on
the reference finite element mesh) are $1.7029188 \times 10^5\ \si{\hertz}$ and
$7.4091162 \times 10^5\ \si{\hertz}$ for the lowest and tenth lowest
frequencies, respectively, at the X-point.  At the M-point, the reference
solution for the lowest frequency is $2.5329233 \times 10^5\ \si{\hertz}$ and
for the tenth lowest frequency is $7.3191350 \times 10^5\ \si{\hertz}$.  In the
extended finite element analyses, the error in frequency reduces with an
increase in the number of degrees of freedom at a rate of approximately $2p$
with all four values of $p$ tested, matching the a priori estimate for finite
elements.  With quadratic finite elements, error reduces at a rate of
approximately $4$.  As with the circular void problem, the quartic spectral
extended finite element solution provides equivalent accuracy with substantially
fewer degrees of freedom as compared to the quadratic finite element solution.
For eight digits of accuracy in the frequency solution, approximately thirty-fold
fewer DOFs are required when using the X-FEM.

\subsection{Two-phase phononic crystal with ring inclusion}

\begin{figure}[t]
  \centering
  \includegraphics[scale=0.969]{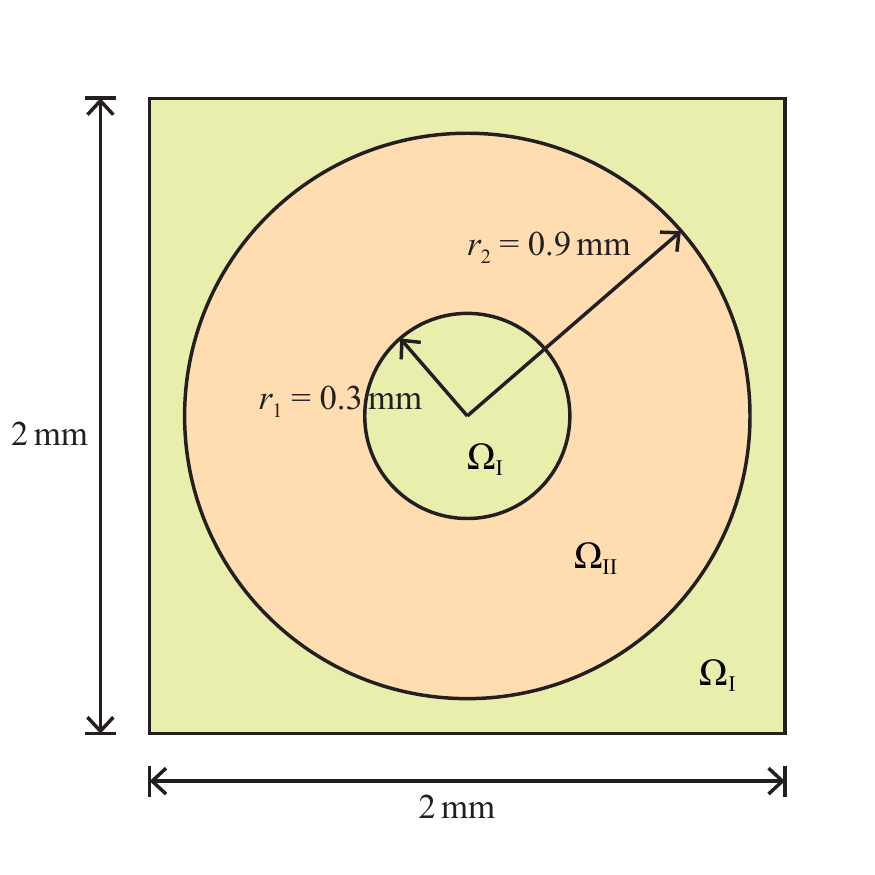}
  \caption{Unit cell geometry of a two-phase phononic crystal with a ring
    inclusion.}
  \label{fig:ex-bimat-nestcircle-setup}
\end{figure}

\begin{figure}
  \centering
  \begin{subfigure}{3.15in}
    \includegraphics[scale=0.969]{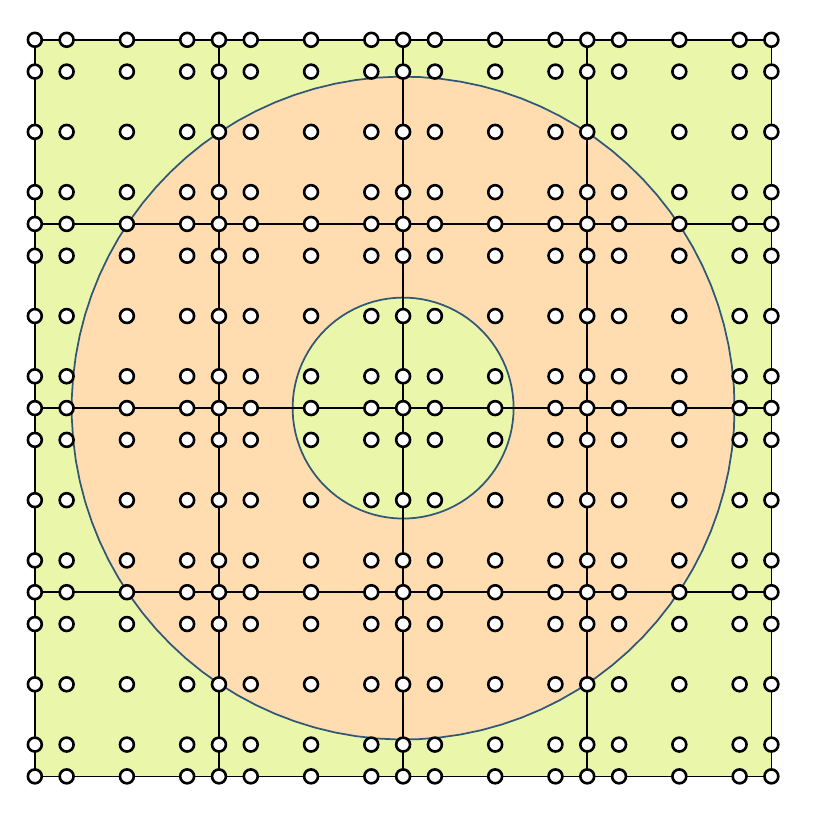}
    \caption{}\label{fig:ex-bimat-nestcircle-mesh-xfem}
  \end{subfigure}
  \begin{subfigure}{3.15in}
    \includegraphics[width=3.15in]{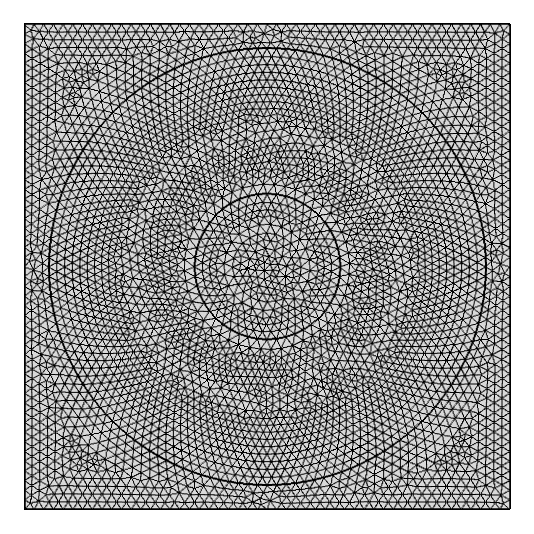}
    \caption{}\label{fig:ex-bimat-nestcircle-mesh-fem}
  \end{subfigure}
  \begin{subfigure}{4.5in}
    \centering
    \includegraphics[scale=0.969]{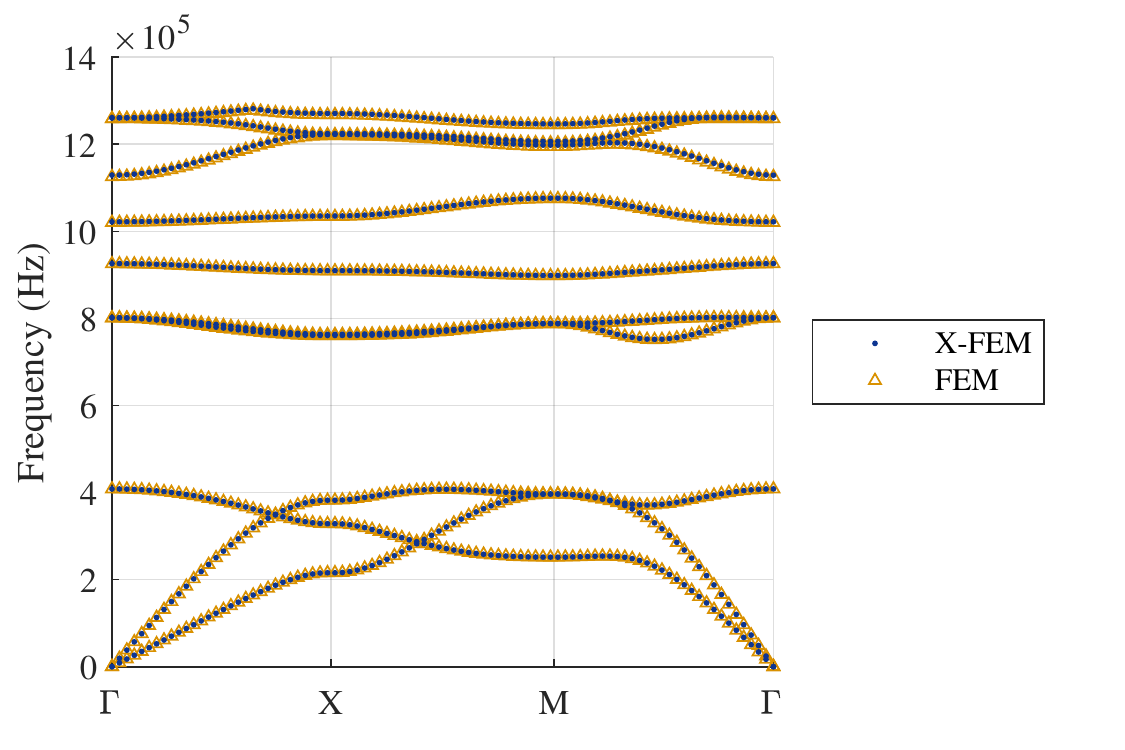}
    \caption{}\label{fig:ex-bimat-nestcircle-bsd}
  \end{subfigure}
  \caption{Band structure diagram for a phononic crystal with a ring inclusion.
    (a) Quartic extended finite element mesh (1,090 total DOFs; all nodes
    contain bimaterial enrichment), (b) quadratic finite element mesh (30,802
    DOFs), and (c) band structure diagram.}
  \label{fig:ex-bimat-nestcircle-bsd}
\end{figure}

As a final example, we generate a band structure diagram for a two-phase
phononic crystal with a ring-shaped inclusion.  The unit cell geometry is
illustrated in \fref{fig:ex-bimat-nestcircle-setup}.  In $\Omega_I$, the
material properties are $E_1 = 2\ \si{\giga\pascal}$, $\nu_1 = 0.3$, and $\rho_1
= 1100\ \si{\kilo\gram\per\meter\cubed}$ and in $\Omega_{II}$, the material
properties are $E_2 = 200\ \si{\giga\pascal}$, $\nu_2 = 0.3$, and $\rho_2 =
3300\ \si{\kilo\gram\per\meter\cubed}$.  The band structure diagram is generated
using both extended finite elements and finite elements.  The meshes used to
generate these solutions are illustrated in
\fref{fig:ex-bimat-nestcircle-mesh-xfem} and
\fref{fig:ex-bimat-nestcircle-mesh-fem}.  With the extended finite element mesh,
16 quartic elements with 1,090 total DOFs are used.  The level set zero is
approximated using cubic Hermite functions, as described in
\sref{ssec:level-set}.  In comparison, the finite element mesh uses 7,600
quadratic triangular elements with 30,802 total DOFs.  The band structure
diagrams generated using the X-FEM and the FEM are displayed in
\fref{fig:ex-bimat-nestcircle-bsd}.  The diagrams match to approximately three
significant digits.  Also of note, a large band gap is present between the third
and fourth frequency with this unit cell configuration and the fourth through
tenth lowest frequencies are nearly constant as the wave vector varies, leading
to additional band gaps.

\section{Conclusions}\label{sec:conclusion}

In this paper, we introduced the spectral extended finite element method (X-FEM)
for constructing band structure diagrams in phononic crystals.  Using the X-FEM
to represent boundaries with spectral finite elements requires accurate and
robust interface reproduction on curved interfaces, accurate and efficient
quadrature over curved interfaces, and an enrichment function over material
interfaces that is capable of reproducing $p$-th order polynomials on both sides
of the interface.  Interface reproduction was accomplished using \bezier{}
curves and approximating level set isocontours with cubic Hermite functions.
Numerical quadrature was conducted using the homogeneous numerical integration
(HNI) method, which represents boundary locations defined by parametric curves
exactly, does not require partitioning elements cut by the interface, reduces
integration to one dimensional integrals without any approximation, and gives
accurate integration that is in fact exact when boundaries are affine or
polynomial curves.  Interfaces are captured through an enrichment function that
is able to reproduce theoretical rates of convergence over curved
interfaces~\cite{Chin:2019:MCI} by accurately capturing the interface location
and by providing linear, ridge-like enrichment on higher order elements.  When
combined, these methods demonstrated accurate band structure diagram
reproduction on two-dimensional domains using minimal degrees of freedom.

The solution efficiency gains possible through spectral extended finite elements
were demonstrated through several examples in \sref{sec:results}.  Band
structure diagrams of phononic crystals with circular voids and circular
inclusions converged much faster with a higher degree of polynomial
reproduction. For example in the circular void problem, to reduce error in
frequency to below $\mathcal{O}(10^{-5})$, 40--50 times more degrees of freedom
were needed with quadratic finite elements as compared to quartic spectral
extended finite elements.  Typically, the utilization of spectral finite
elements is limited by issues of mesh generation and open questions of
polynomial reproduction with non-affine transformations of spectral elements.
However, the framework of the spectral extended finite element method introduced
in this paper eliminates these issues.  Meshes are no longer required to conform
to the locations of voids and material inclusions in phononic crystals;
therefore, elements can remain aligned with the (affine) unit cell structure
ensuring that the higher-order polynomial space is retained.

Eliminating the need for mesh generation and providing efficient generation of
phononic band structure diagrams enables a more rapid understanding of the
properties of various phononic crystals, and speeds up iterative design
processes, including parametric studies and the design of improved phononic
crystals using shape optimization and topology optimization.  For the design and
optimization of phononic crystals, level set representations of geometry are
appealing since they allow for arbitrary smooth, curved boundaries.  Level set
methods have already been successfully applied to topology optimization and
shape optimization~\cite{Osher:2001:LSM,Wang:2003:ALS,Allaire:2004:SOU},
demonstrating the appeal of combining these ideas.  For these applications, the
use of the spectral X-FEM presented in this paper would be worth pursuing.

\section*{Declaration of competing interest}
The authors declare that they have no known competing financial interests or
personal relationships that could have appeared to influence the work reported
in this paper.

\section*{Acknowledgements}
E.B.\ Chin and N.\ Sukumar gratefully acknowledge the research support of Sandia
National Laboratories to the University of California, Davis.  Additional
financial support to E.B.\ Chin from the ARCS Foundation Northern California is
also acknowledged. This work was performed under the auspices of the U.S.
Department of Energy by Lawrence Livermore National Laboratory under Contract
DE-AC52-07NA27344. Ankit Srivastava acknowledges support from the NSF CAREER
grant \#1554033 to the Illinois Institute of Technology and NSF grant \#1825354
to the Illinois Institute of Technology.

\section*{Disclaimer}
{\small
This document was prepared as an account of work sponsored by an agency of the
United States government. Neither the United States government nor Lawrence
Livermore National Security, LLC, nor any of their employees makes any warranty,
expressed or implied, or assumes any legal liability or responsibility for the
accuracy, completeness, or usefulness of any information, apparatus, product, or
process disclosed, or represents that its use would not infringe privately owned
rights. Reference herein to any specific commercial product, process, or service
by trade name, trademark, manufacturer, or otherwise does not necessarily
constitute or imply its endorsement, recommendation, or favoring by the United
States government or Lawrence Livermore National Security, LLC. The views and
opinions of authors expressed herein do not necessarily state or reflect those
of the United States government or Lawrence Livermore National Security, LLC,
and shall not be used for advertising or product endorsement purposes.
}

\bibliography{references}

\end{document}